\documentclass[twocolumn]{aastex62}
\usepackage{amssymb,amsmath,graphicx,natbib,hyperref}
\usepackage{color}
\usepackage{enumerate}
\usepackage{longtable}

\usepackage[encapsulated]{CJK}
\usepackage[utf8x]{inputenc}
\DeclareGraphicsExtensions{.pdf,.PDF,.png,.PNG,.jpg,.JPG,.jpeg,.JPEG}
\newcommand{\hi}{{\rm H}{\textsc i}}
\graphicspath{}

\shorttitle{\hi-SF correlation}
\shortauthors{Zhou et al.}

\begin{document}
\begin{CJK*}{UTF8}{gbsn}

\title{The relation between \hi\ Gas and Star Formation Properties in Nearby Galaxies}
\author{Zhimin Zhou}
\affiliation{Key Laboratory of Optical Astronomy, National Astronomical Observatories, Chinese Academy of Sciences, Beijing, 100012, China}
\correspondingauthor{Zhimin Zhou}
\email{zmzhou@bao.ac.cn}

\author{Hong Wu}
\affiliation{Key Laboratory of Optical Astronomy, National Astronomical Observatories, Chinese Academy of Sciences, Beijing, 100012, China}
\affiliation{College of Astronomy and Space Sciences, University of Chinese Academy of Sciences, Beijing 100049, China}

\author{Xu Zhou}
\affiliation{Key Laboratory of Optical Astronomy, National Astronomical Observatories, Chinese Academy of Sciences, Beijing, 100012, China}

\author{Jun Ma}
\affiliation{Key Laboratory of Optical Astronomy, National Astronomical Observatories, Chinese Academy of Sciences, Beijing, 100012, China}
\affiliation{College of Astronomy and Space Sciences, University of Chinese Academy of Sciences, Beijing 100049, China}

\begin{abstract}
In this paper, we present some correlations of neutral hydrogen (\hi) gas and physical properties of galaxies to investigate the role of atomic gas in governing galaxy evolution. We build a \hi-detected sample including 70 galaxies that are detected by ALFALFA in a 15 deg$^2$ region, and derive their star formation distribution based on the images of $\rm H\alpha$ narrow-band observed here. In general, \hi-detected galaxies have low surface density of stellar mass and active star formation. Additionally, most of the galaxies are in good agreement with the star-forming main sequence, consistent with the previous findings. We confirm the dependence of star formation (SF) in galaxies on \hi\ gas at least on global scale, i.e., star formation rate (SFR) generally increases with \hi\ mass, specific star formation rate (SSFR$\equiv$SFR/M$_*$) increases with \hi\ fraction ($f_{\hi}$) even for a given stellar mass, and \hi-based star formation efficiency (SFE) mildly increases with the stellar mass and SFR surface density. Based on the distribution of stellar mass and star formation, we calculate the morphology indices of the sample, and analyze the dependence of $f_{\hi}$ and SFE on them. The weak correlations between SFE and morphological indexes imply a weak physical link between \hi\ and star formation in small scale. We find that $f_{\hi}$ mildly increases with the asymmetry and decreases with the concentration of galaxies, suggesting that the \hi\ gas supply and its effect are likely correlated with external processes in the extended disks of galaxies. 
\end{abstract}

\section{Introduction}

\label{sec:intro}

The interstellar medium (ISM) is the fuel for the star formation and accretion processes in galaxies, and is fundamental to galaxy formation and evolution. The conversion of gas into stars is the basic process which governs the life of a galaxy. As an important cold ISM component, atomic hydrogen, is widespread in galaxy disks and can be traced by the 21-cm line of neutral hydrogen (\hi).  \hi\ gas is able to convert to molecular gas $\rm H_2$ in regions where \hi\ reaches high enough densities and metallicities to become shielded from interstellar ultraviolet (UV) photons \citep{Krumholz2012}.  Because molecular gas is directly related to star formation, atomic gas can be regarded as the reservoir of future star formation in a galaxy \citep{Kauffmann2015}. Thus, investigating how and why the atomic gas content of a galaxy varies with stellar mass and other physical properties can provide initial clues to its growth history.

There have been many works that present the relationships of \hi\ content with galaxy properties, such as stellar mass, color, morphology, star formation, and environment \citep[e.g.,][]{Fumagalli2008, Cortese2011, Wang2011, Hughes2013}. In general, galaxies with more gas are bluer and more actively star forming.
Observations have shown the well-known connection between the star formation rate (SFR) and gas content as parameterized by the Kennicutt-Schmidt Law \citep[K-S Law;][]{Kennicutt1998a, Kennicutt2012}, indicating that integrated SFR surface density is regulated by the gaseous surface density. Motivated by the K-S star formation law, many investigations have explored the scaling relations of the \hi\  gas fraction with galaxy structure and properties, and calibrated a series of photometric estimators of \hi\ mass fraction with stellar mass, stellar surface mass density, colors, and specific SFR \citep[SSFR$\equiv$SFR/M$_*$; e.g.,][]{Zhang2009, Catinella2010, Li2012}. \citet{Huang2012} investigate the global scaling relations and fundamental planes linking stars and gas for a sample of \hi-bearing galaxies. They found that 96\% of their sample belong to the blue cloud, with the average gas fraction f$_{HI}$ $\equiv M_{HI}/M_{*} \sim 1.5 $.
\citet{Bothwell2013} found a three-dimensional fundamental relation between stellar mass, gas-phase metallicity, and \hi\ mass, with \hi-rich galaxies being more metal poor at a given stellar mass, which is likely more fundamental than the relation between metallicity, SFR and mass.

Blind \hi\ surveys in recent years have produced large, unbiased samples of galaxies selected by \hi\ mass, providing invaluable insights into the research of \hi\ gas in nearby galaxies \citep[see][for a review]{Giovanelli2016}, such as the \hi\ Parkes All-Sky Survey \citep[HIPASS;][]{Barnes2001, Meyer2004} and the \hi\ Jodrell All-Sky Survey \citep[HIJASS;][]{Lang2003}. The Arecibo Legacy Fast ALFA (ALFALFA) survey, a wide-field \hi\ survey completed most recently, maps 7000 deg$^2$, detects more than 30,000 extragalactic \hi\ sources, and provides the first full census of extragalactic \hi\ line sources over a cosmologically significant volume in the local universe out to $z\sim$0.06 \citep{Giovanelli2005, Haynes2011}. Based on ALFALFA catalog, many samples are selected and analyzed in different galaxy regimes. For example, the HIghMass sample selected by \citet{Huang2014} includes \hi-massive and high-\hi-fraction galaxies, the Survey of \hi\ in Extremely Low-mass Dwarfs \citep[SHIELD;][]{Cannon2011} includes a sample of galaxies with \hi\ masses below $10^7$ $M_{\odot}$ outside the Local Group. Besides that, \citet{Sistine2016} utilizes a large sample of \hi-selected galaxies between 20$-$100 Mpc from the ALFALFA survey to study star formation in the local universe based on the narrow-band imaging of $\rm H\alpha$ emission line. \citet{Gavazzi2012} completed a similar $\rm H\alpha$ narrow-band imaging survey of an \hi\ line flux-selected sample of Local Supercluster galaxies to probe the role of the environment in shaping the star formation properties of galaxies.

Most of the target selections in previous studies usually focused on certain types of galaxies with specific criteria of physical properties \citep{Wang2017a}, and missed a fraction of galaxies which are also detected in \hi\ surveys. Even in the $\alpha$.40-SDSS-GALEX sample of \citet{Huang2012}, there are nearly one-third of galaxies excluded from the parent sample.
Therefore, a complete, systematic sample of nearby galaxies is urgently needed to help investigate scaling relations between \hi\ gas and galactic properties, determine the differences in the physical properties of \hi-rich and -poor galaxies.

\begin{figure}
	\centering
	\includegraphics[width=\columnwidth]{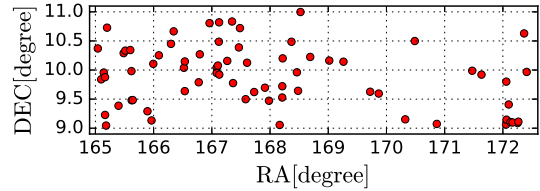}
	\caption{Sky distribution of our sample. The sample includes all of the galaxies detected by ALFALFA survey in the region $\rm {11^h \leq R.A. \leq 11^h30^m, 9^\circ \leq decl. \leq 11^\circ}$.}
	\label{fig_sky}
\end{figure}

\begin{figure*}
	\centering
	\includegraphics[width=0.4\hsize]{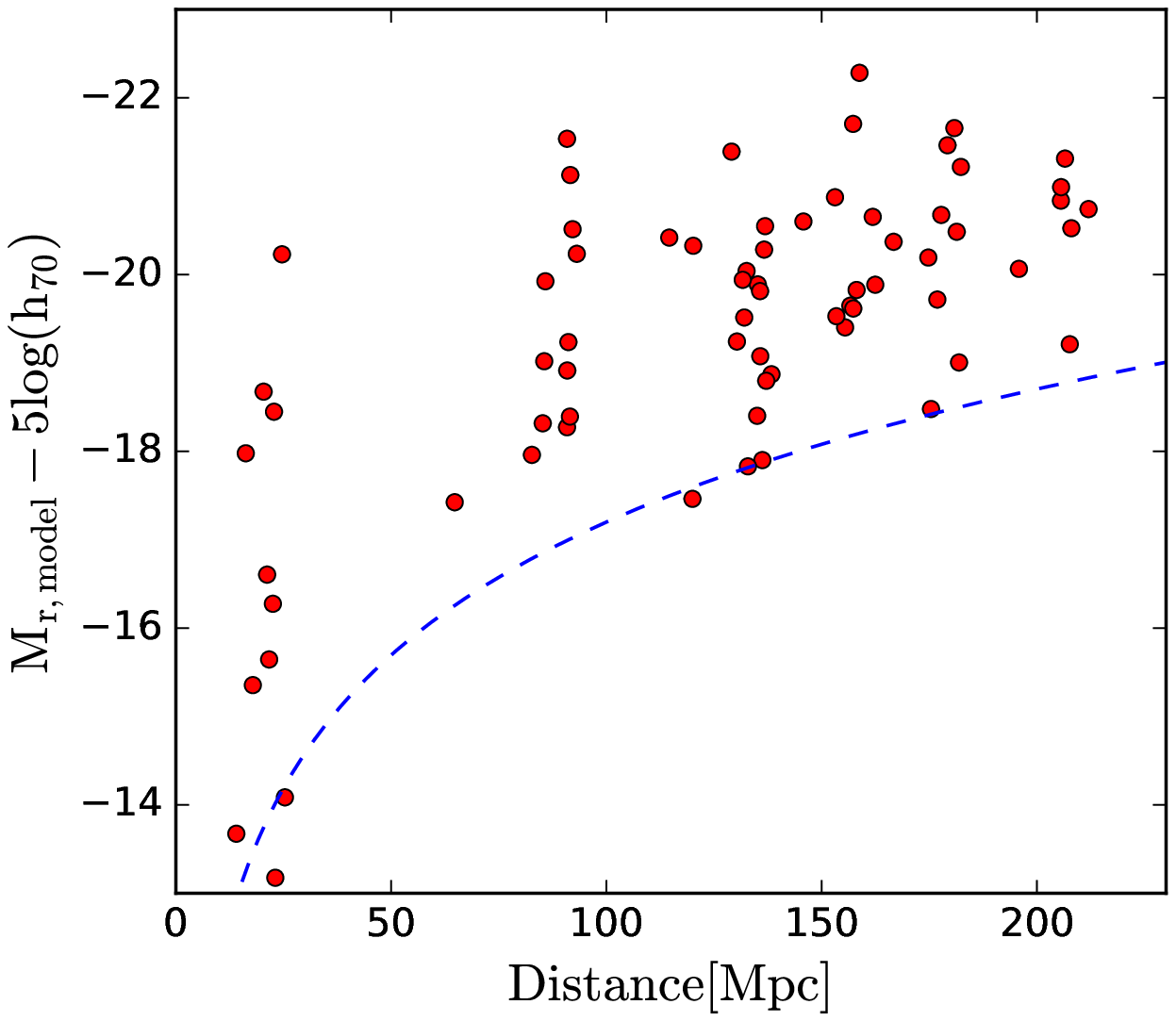}
	\includegraphics[width=0.4\hsize]{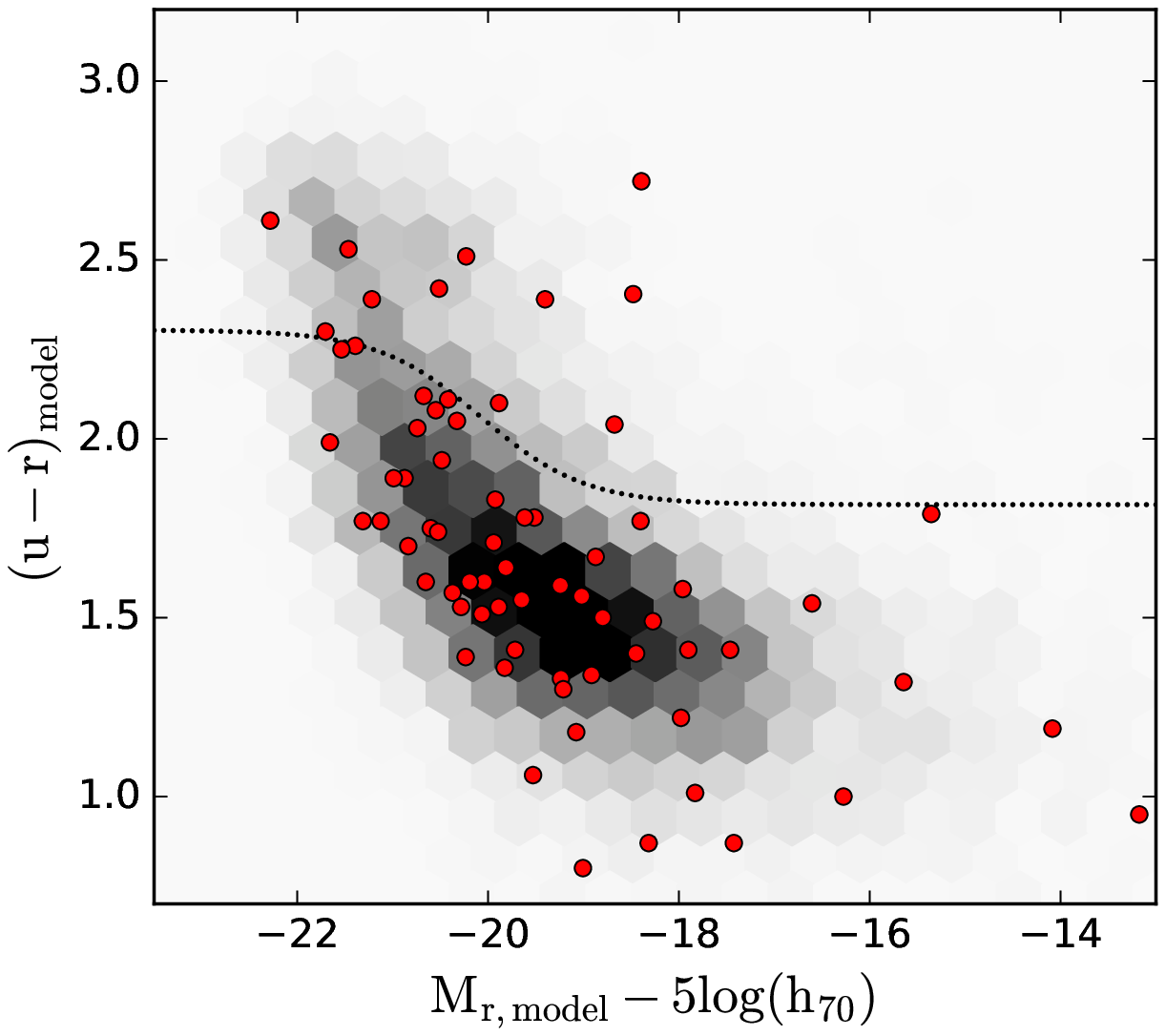}
	\caption{Distribution of our samples in the distance vs. magnitude plane (left panel) and color-magnitude diagram (right panel). The blue curve in the left panel shows the observed flux limits at $m_r$=17.77 magnitude for SDSS legacy galaxy redshift galaxies. In the right panel, the superposed dashed line is the optimum divider separating the red sequence from the blue cloud from \citet{Baldry2004}, the gray background depicts the distribution for the ALFALFA-SDSS overlap galaxies shown in \citet{Haynes2011}.}
	\label{fig_dist}
\end{figure*}

\begin{figure*}
	\centering
	\includegraphics[width=0.8\hsize]{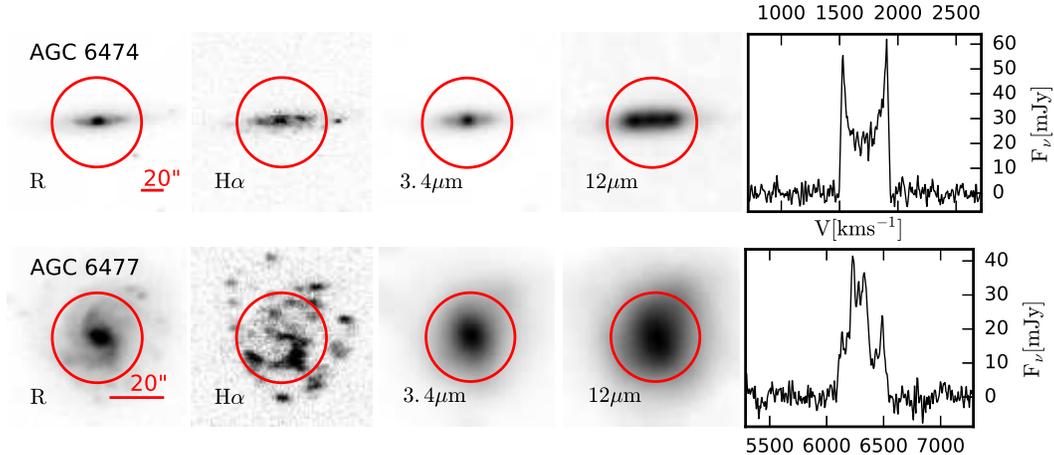}
	\caption{Images and \hi\ spectrum of AGC 6474 ($\it top$) and AGC 6477 ($\it bottom$). The images of broad-band {\it R}, continuum-subtracted narrow-band H$\alpha$, WISE IR 3.4$\mu$m and 12$\mu$m, along with the \hi\ Spectrum from ALFALFA are shown from left to right panels, respectively. The galaxy names are specified in the panels. North is up and east is left along with a 20$''$ scale bar in the first image for each object. Petrosian radii of SDSS $r$ band are overlaid on the images.}
	\label{fig_image}
\end{figure*}

In this paper, we try to present more about the relations between cold gas and star formation in nearby galaxies. We focus on a 15 deg$^2$ region of ALFALFA survey, select all galaxies detected by ALFALFA in this sky area, take the $\rm H\alpha$ narrow-band imaging to trace their star formation activity, and analyze how the star forming properties correlate with \hi\ gas.

The paper is organized as follows. In Section \ref{sec:data}, we introduce in detail the sample selection, the data used in this paper, $\rm H\alpha$ narrow-band imaging, and data reduction. In Section \ref{sec:method} we describe our methodology for measurement the SFR and stellar mass ($M_*$) of \hi\ sample. Section \ref{sec:analysis} presents the  physical properties of our sample, and analyzes the relations between the \hi\ content, star formation, stellar mass, and galactic morphology. We discuss our findings in Section \ref{sec:discuss} and summarize the results in Section \ref{sec:summary}.





In the following, we adopt a standard $\Lambda$CDM cosmology with ${H}_0 = 70 {\rm~km~s^{-1}~Mpc^{-1}}$, $\rm \Omega_m = 0.3$, and $\Omega_{\Lambda} = 0.7$.

\section{Data and Sample}
\label{sec:data}

\subsection{Samples and ALFALFA data}

The ALFALFA survey presents a database of 21 cm \hi\ line sources over $\sim$ 7000 deg$^2$ of the sky at high Galactic latitude, yielding a source density of 5.3 galaxies per deg$^2$. This survey covers the recessional velocity range up to 18,000 km s$^{-1}$ with the resolution of $\sim$5 km s$^{-1}$, and can detect \hi\ masses as low as 10$^6$ $M_{\odot}$ and as high as 10$^{10.8}$ $M_{\odot}$ with positional accuracies typically better than 20$''$ \citep{Haynes2011}. In the ALFALFA catalog, the detections with signal-to-noise ratio S/N$>$6.5 are categorized as ``Code 1'' which are nearly 100\% reliable, the sources categorized as ``Code 2'' have low S/N ($<6.5$) but coincide with likely optical counterparts at the same redshift. The \hi\ mass $M_{\hi}$ is computed via the formula $M_{\hi} = 2.356 \times 10^5 D^2_{\rm Mpc} S_{21}$, where $D_{\rm Mpc}$ is the distance to the galaxy in the unit of Mpc, and $S_{21}$ is the integrated \hi\ line flux density of the source, in Jy km s$^{-1}$.

Our sample is selected from a $\sim$7.5 $\times$ 2 deg$^2$ sky region of the ALFALFA survey ($\rm 11^h \leq R.A. \leq 11^h30^m$, $\rm 9^\circ \leq decl. \leq 11^\circ$). It contains all detections with both the code 1 and 2, and adopts the \hi\ measures, distances, and \hi\ masses presented in the $\alpha$.40 catalog of \citet{Haynes2011}. There are 70 extragalactic \hi\ sources in our sample, 58 with code 1 and 12 with code 2.


Figure \ref{fig_sky} presents the distribution of the sample galaxies in the sky region. This region is located in the field near the Virgo cluster, and most of our samples are field galaxies. It should be noted that \hi\ galaxies used in this study are only the objects that can be detected by ALFALFA survey, and the \hi-undetected galaxies in the same region may also contain some \hi\ gas, but under the detection limiting of ALFALFA \citep{Haynes2011}. Beside that, based on the results from \citet{Gavazzi2013a}, there are no clear difference in the physical properties for galaxies with projected distance $>$ 10$^\circ$ from M 87. Thus, although our sample is located in the sky region near the Virgo cluster, the effect of coherent structures from Virgo cluster on the property of our galaxies can be negligible.

As shown in Figure \ref{fig_dist}, most of the galaxies in our sample are located at the distance of 20-230 Mpc with the median distance of $\sim$ 135 Mpc. The right panel of Figure \ref{fig_dist} illustrates the distribution of the samples in the color-magnitude diagram with the gray background for all of the ALFALFA-SDSS overlap galaxies shown in \citet{Haynes2011}. We also superpose the optimum divider used by \citet{Baldry2004} to separate the red sequence from the blue cloud. \hi-detected galaxies are clearly dominated by blue spiral galaxies.

In Figure \ref{fig_imageshow}, we present the \hi\ profiles of all galaxies, which are derived from ALFALFA. Nearly a half of the sample show the typical two-horned profile of a spiral galaxy, especially for the edge-on galaxies, such as AGC 6474. Besides that, about 10 galaxies have a single peak of the \hi\ profile with velocity width $W50 \leq 50~\rm{km\ s^{-1}}$, such as AGC 213090 and AGC 6438, which indicates galaxies of sufficiently low mass \citep{Courtois2009}. AGC 6248 is the most anorexic of galaxies in the our sample, its $W50$ is $26~\rm km\ s^{-1}$, and the stellar mass is only $6.3 \times 10^6~M_{\odot}$. There are also some anomalous behaviors in the \hi\ profiles, such as multi-peaks (e.g., AGC 6477, AGC 215241), asymmetric (e.g., AGC 6209, AGC 215236). If true, they are likely derived by the physical process in the galaxy evolution, while some may be due to low signal-to-noise S/N or contamination from near neighbors.

\subsection{$\rm H\alpha$ narrow-band observation and data reduction}

The H$\alpha$ emission in galaxies can be used as a good star formation tracer \citep{Kennicutt1998b, Zhu2008}. We performed H$\alpha$ narrow-band observations for the \hi-detected galaxies in the spring of 2012. The observations were completed by two instruments on two telescopes, respectively. One is the BAO Faint Object Spectrograph and Camera (BFOSC) mounted on the 2.16m telescope \citep{Fan2016} at Xinglong Observatory of the National Astronomical Observatories, Chinese Academy of Sciences (NAOC). The other is the Yunnan Faint Object Spectrograph and Camera (YFOSC) mounted on 2.4m telescope \citep{Fan2015} in the Lijiang observational station of Yunnan observatories, Chinese Academy of Sciences (YNAO). Most objects were observed in photometric conditions: six nights in January and March for BFOSC, and six nights in April for YFOSC.

In the 70 \hi-detected galaxies, 43 galaxies were imaged by BFOSC, which has a Lick 1242 $\times$ 1152 CCD detector with the pixel scale of 0\farcs457 pixel$^{-1}$ and field of view (FOV) of $9\farcs46 \times8\farcs77$. The others were observed with YFOSC, whose CCD has the FOV of $9\farcs6 \times 9\farcs6$ with the pixel scale of 0\farcs283 pixel$^{-1}$. The galaxies were imaged with the broad-band $\it R$ filter and narrow-band filters covering H$\alpha$ emission line (rest-frame wavelength of $\lambda6563\AA$). We used the same set of H$\alpha$ filters in BFOSC and YFOSC, which includes 11 narrow-band filters with the central wavelengths between 656.2 and 706.0 nm with an FWHM of 7 nm, corresponding to redshifts of $0-22000$ $\rm km\ s^{-1}$. It should be noted that the H$\alpha$ narrow-band filters also include two [N {\sc ii}] $\lambda\lambda$6548, 6583 emission lines, a proper correction for [N {\sc ii}] emission is required before the final H$\alpha$ flux is computed. The total exposure times were typically $300-600$ s for $\it R$ and $600-1800$ s for H$\alpha$ based on the observational condition.

We took our data reduction using IRAF\footnote{IRAF is the Image Reduction and Analysis Facility written and supported by the IRAF programming group at the National Optical Astronomy Observatories (NOAO) in Tucson, Arizona, which is operated by AURA, Inc. under cooperative agreement with the National Science Foundation}, and created continuum-subtracted images following the procedure described in \citet{Zhou2015}. We summarize this procedure below. First of all, our images were reduced using the standard reduction pipeline, including subtracting overscan and bias, correcting bad pixels and flat fielding, removing cosmic rays, and adding astrometric solutions. Then, flux calibration of broad-band images was made using the field stars cross-matched with SDSS DR10 catalog, whose magnitudes are converted to $UBVRI$ system following on Robert Lupton's transformation equations on SDSS website\footnote{http://www.sdss3.org/dr10/algorithms/sdssUBVRITransform.php}. Next, the scaled $R$-band image as stellar continuum was subtracted from each H$\alpha$ image, and the scale factor for each image was calculated with the flux of a dozen unsaturated field stars near the galaxy. Finally, flux calibration was made for H$\alpha$ images based on the scale factors and effective transmissions of narrow-band and $R$-band filters.

For the H$\alpha$ images, we then made some corrections before we got the final integrated H$\alpha$ flux. First, a small fraction ($\sim$ 6\%) of H$\alpha$ emission is lost in the process of continuum removal \citep{Kennicutt2008}, so we recovered this flux by comparing the scale factors and effective transmissions of narrow-band and $R$ filters. Second, we corrected the Galactic foreground extinction based on the \citet{Schlegel1998} dust maps and the assumption of $R_V = 3.1$ absorbing medium. Then, the deblending from [N {\sc ii}] was corrected based on the Equation (B1) of \citet{Kennicutt2008}. Finally, the uncertainties of our integrated H$\alpha$ fluxes are $\sim$ 10\%.

\subsection{Ancillary data}
To significantly improve the statistics of our samples, a few more data sets are used for analysis or comparison in this paper.

We supplement the photometric data from {\it the Wide Field Infrared Survey Explorer} \citep[{\it WISE};][]{Wright2010} to measure SFRs and stellar masses of our \hi-detected sample. WISE has mapped the entire sky at 3.4, 4.6, 12, and 22$\mu$m with an angular resolution of 6\farcs1, 6\farcs4, 6\farcs5, and 12\farcs0, respectively. We searched the ALLWISE database for detections within 6$''$ of the optical coordinates of \hi-detected galaxies, and derived their profile-fit photometry magnitudes in these infrared bands.

The Sloan Digital Sky Survey \citep[SDSS;][]{York2000} provides homogeneous photometric and spectroscopic data of high quality for very large and objectively selected samples of galaxies over one-third of the sky. We derived galaxy properties from the MPA-JHU spectroscopic analysis which presents catalogs of the measurements of absorption line indices and emission line fluxes as well as stellar masses, SFRs, and oxygen abundance for the SDSS-DR7 legacy galaxy redshift sample \citep{Brinchmann2004, Kauffmann2003}.

Figure \ref{fig_image} shows our images and \hi\ spectrum used for two objects, including continuum-subtracted H$\alpha$, {\it R}-broadband, WISE 3.4$\mu$m and 12$\mu$m images. The data for all of the sample can be found in Figure \ref{fig_imageshow}.

\section{MEASUREMENTS}
\label{sec:method}

\begin{figure}
	\centering
	\includegraphics[width=0.8\columnwidth]{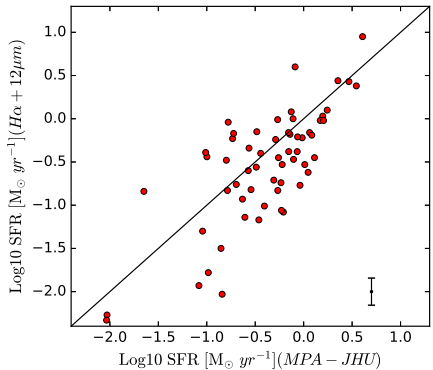}
	\caption{Comparison of H$\alpha + 12~\mu$m based SFR (this work) with MPA-JHU SFRs from SDSS spectroscopic analysis. The solid line marks the one-to-one relation. The error bar indicates the average error of our SFR calculation.}
	\label{fig_sfr}
\end{figure}
\begin{figure}
	\centering
	\includegraphics[width=0.8\columnwidth]{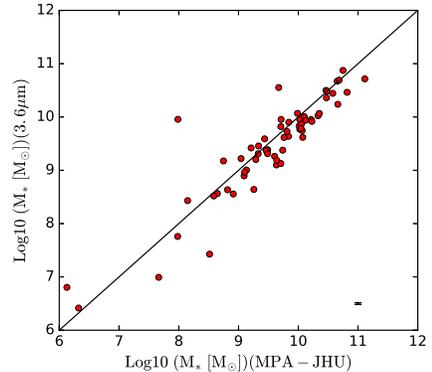}
	\caption{Comparison of WISE 3.4$\mu$m based stellar mass (this work) with MPA-JHU results. The solid line marks the one-to-one relation. The error bar indicates the average error of the calculation.}
	\label{fig_mass}
\end{figure}

\begin{figure*}
	\centering
	\includegraphics[width=0.8\hsize]{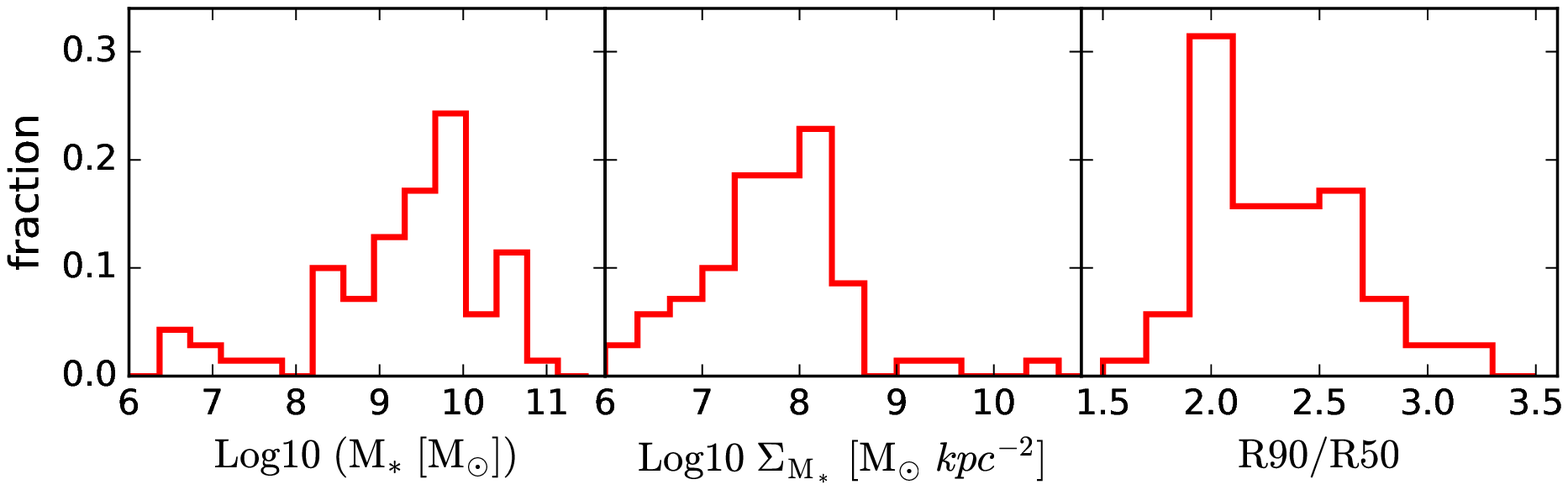}
	\includegraphics[width=0.8\hsize]{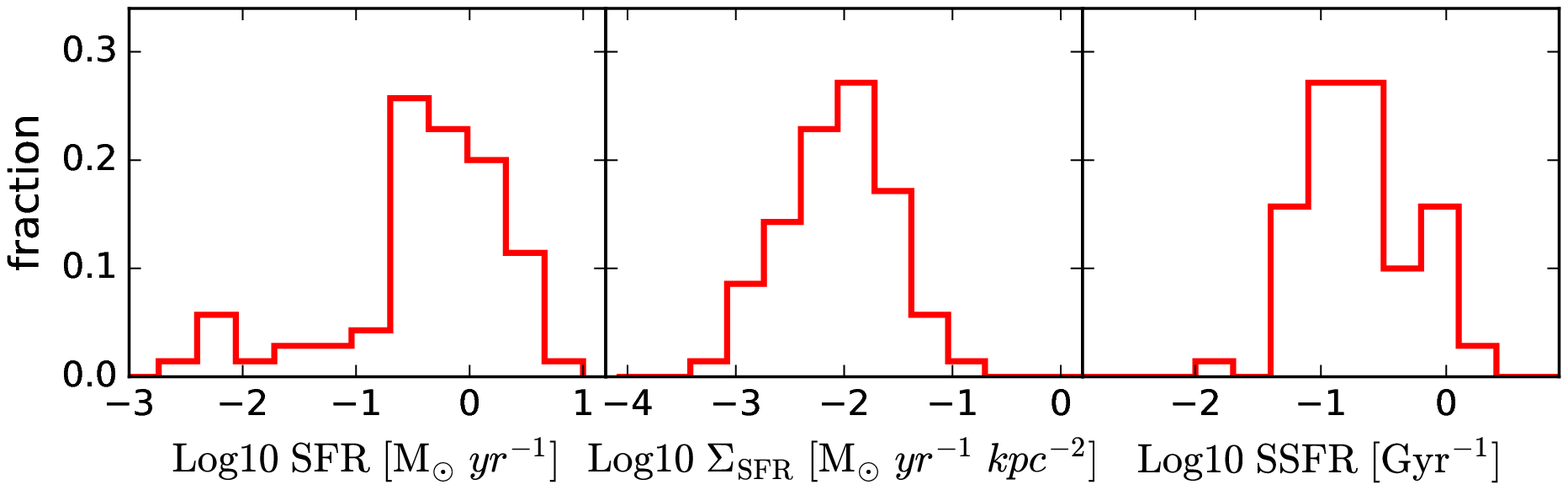}
	\caption{Distribution of the basic properties for the sample. The histograms show the stellar mass $M_*$, SFR, SSFR, surface densities of stellar mass and SFR, concentration index $R90/R50$ for galaxies in our \hi-detected galaxies.}
	\label{fig_hist}
\end{figure*}

\begin{figure*}
	\centering
	\includegraphics[width=0.4\hsize]{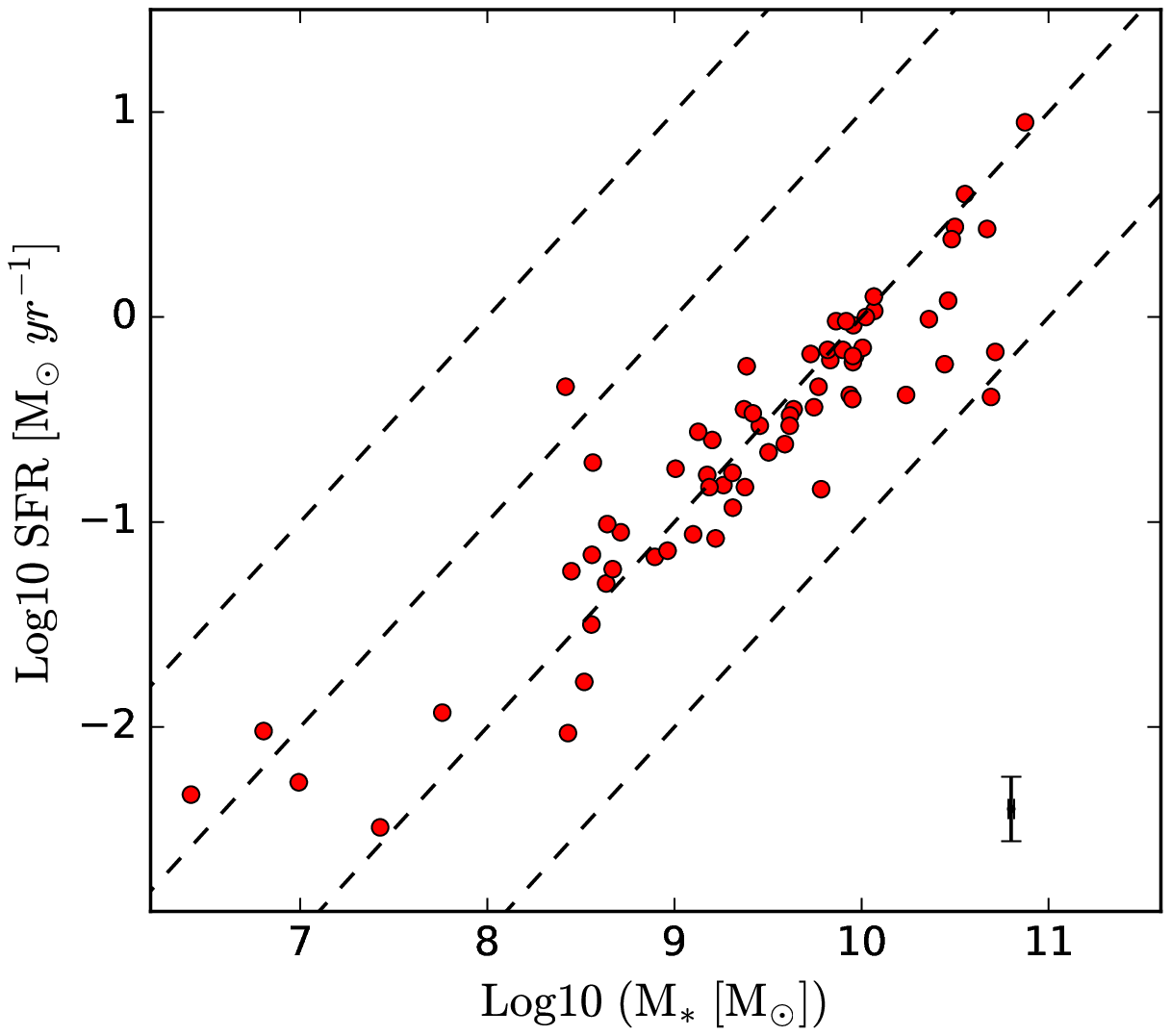}
	\includegraphics[width=0.4\hsize]{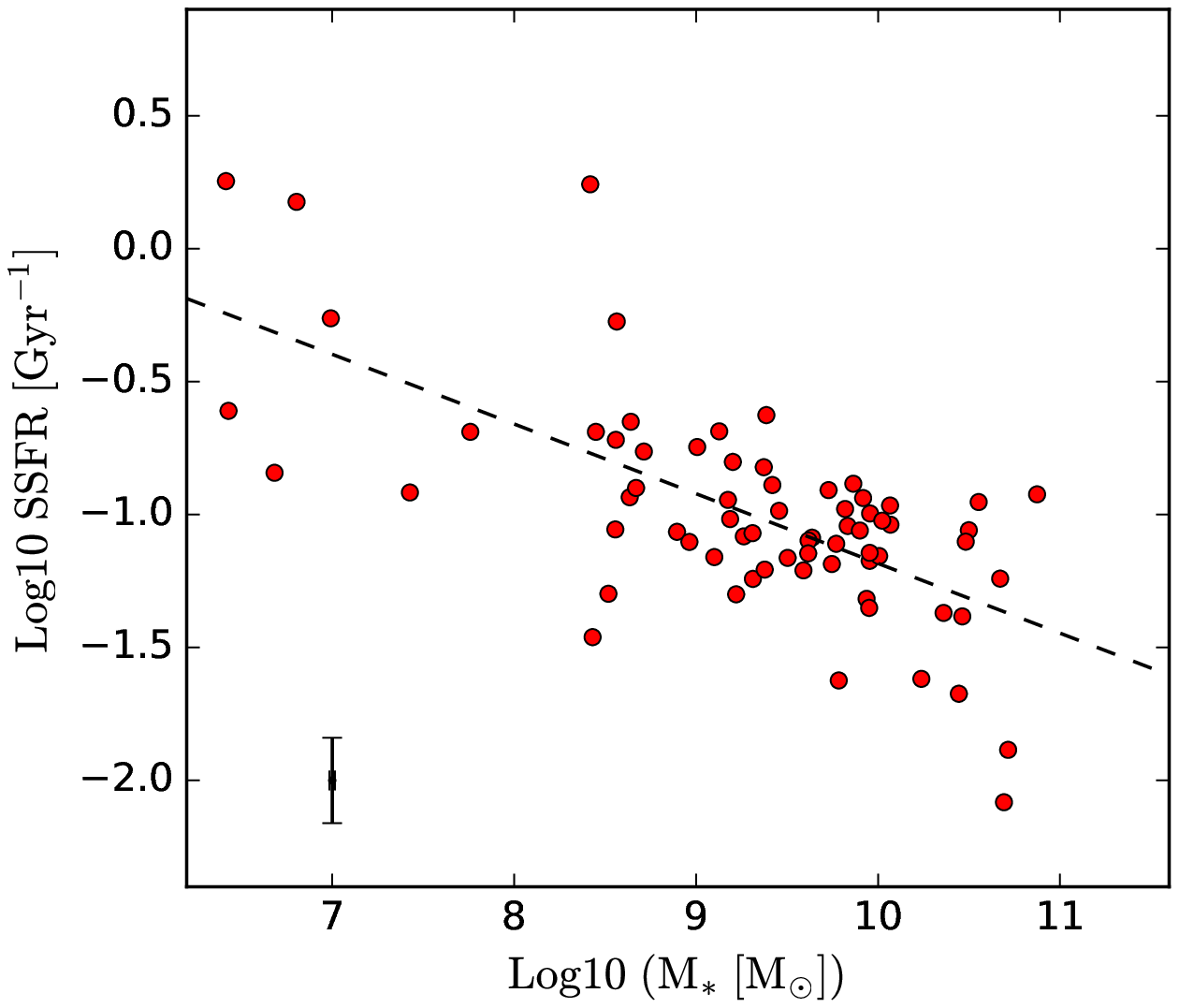}
	\caption{Distributions of our samples in the SFR-$M_*$ (left panel) and SSFR-$M_*$ (right panel) planes. The dashed lines in the left panel mark the positions for SSFR = $10^1$ to $10^{-2}$ $\rm Gyr^{-1}$, top to bottom, with steps of one order of magnitude. In the right panel, the dashed line represents the linear fit for the \hi-detected sample.}
	\label{fig_SFR_GM}
\end{figure*}

We calculate SFRs of the galaxies using the H$\alpha$ emission and WISE 12$\mu$m photometric fluxes with the method of \citet[][their Equation 14]{Wen2014}:
\begin{eqnarray}
	{\rm SFR}(M_{\odot}~{\rm yr}^{-1})=10^{-41.270} \times [L({\rm H\alpha})_{\rm obs}
		\nonumber\\
	+0.038L(12\mu m)]({\rm erg~s^{-1}}).
\end{eqnarray}
This formula is calibrated with solar abundance and the Salpeter initial mass function (IMF) with a slope of 2.35 for stellar masses in the range 0.1--100 $M_{\odot}$. Given the uncertainties of our photometric results and SFR calibration we used, the median variation of the SFRs we calculated is $\sim$ 0.033 dex along with most of them ($>$ 80\%) less than 0.1 dex.

There are 62 \hi-selected galaxies in SDSS MPA-JHU catalog. We compared our result with the global SFRs from the MPA-JHU catalog in Figure \ref{fig_sfr}. The latter is derived based on fiber spectroscopic analysis and aperture correction \citep{Brinchmann2004}. Our SFRs agree well with MPA-JHU SFRs, with no systematic offset and a median difference of 0.03 dex, although noticeable outliers exist. There are some discrepancies between the two SFRs for several galaxies. Large deviations for the outliers are likely due to the differences between the real stellar distribution and models used for aperture correction in \citet{Brinchmann2004}.

The stellar mass of our \hi-selected galaxies is calculated using WISE 3.4 $\mu$m luminosities with the equation in \citet[][their Equation 2]{Wen2013}:
\begin{equation}
	{\rm Log_{10}}{M(M_{\odot})}=-0.04+1.12 {\rm Log_{10}}\nu L_{\nu}{[3.4\mu m]}(L_{\odot}).
\end{equation}

We compared our results with those derived from SDSS MPA-JHU catalog in Figure \ref{fig_mass}. MPA-JHU stellar masses are obtained from fits to the photometry with BC03 models \citep{Bruzual2003} spanning a large range in star formation histories \citep{Kauffmann2003}. The stellar masses estimated with the two methods are in excellent agreement for most of the galaxies with a median difference of 0.13$\pm$0.28 dex.

Measurements and properties for the 70 galaxies are presented in Table \ref{table:obsevations} and \ref{table:properties}. The first table includes the information of the galaxies and the fluxes of \hi\, $\rm H\alpha$ and WISE IR emission. The second one contains the physical properties and morphological parameters for galaxies.

\section{Analysis}
\label{sec:analysis}

\subsection{Physical properties of \hi-detected galaxies}

Figure \ref{fig_hist} presents the distribution of physical properties of the \hi-detected galaxies. The histograms show stellar mass $M_*$, SFR, SSFR, surface densities of stellar mass and SFR, concentration index. The concentration index is traced by the size ratio, $R_{90}/R_{50}$, which are the semi-major axes of the ellipses that encompasses 90\% and 50\% of the SDSS {\it r}-band Petrosian flux, respectively. The stellar mass surface density is given by $\Sigma_{M_*} = M_* / (2\pi R_{50}^2)$, and the  SFR surface density is given by $\Sigma_{\rm SFR} = {\rm SFR}/(2\pi R_{50}^2)$. The mean stellar mass is ${\rm log}_{10} M_*$ $=$ 9.48 for the galaxies, and the mean of the stellar mass surface density $\Sigma_{M_*}$ is ${\rm log_{10}}\Sigma_{M_*} = 7.69$. In general, the galaxies in our sample have active star formation with $\rm <log_{10} SFR> = -0.28$ and $\rm <log_{10} SSFR> = 0.71$.

To explore further the dependence of atomic gas on integrated galaxy properties, the distributions of the samples in the SFR-$M_*$ and SSFR-$M_*$ planes are presented in Figure \ref{fig_SFR_GM}. Clearly, the sample of \hi-detected galaxies is located following the star-forming main sequence which shows a linear correlation between the logarithm of the SFR and the logarithm of $M_*$ \citep{Brinchmann2004,Elbaz2007,Lee2015}. These galaxies have SSFRs in the range of $0.1-1$ Gyr$^{-1}$, consistent with what we found in Figure \ref{fig_hist}.

For the galaxies in SSFR-$M_*$ plane of Figure \ref{fig_SFR_GM}, we find that SSFRs decrease with increasing stellar mass. We perform robust linear fittings to their relations, and find that the slope is $\sim -0.33$ for all of the \hi-detected sample. The slope is much shallower than that for the relation of non-star-forming galaxies \citep{Bauer2013}.

\begin{figure*}
	\centering
	\includegraphics[width=1.1\hsize]{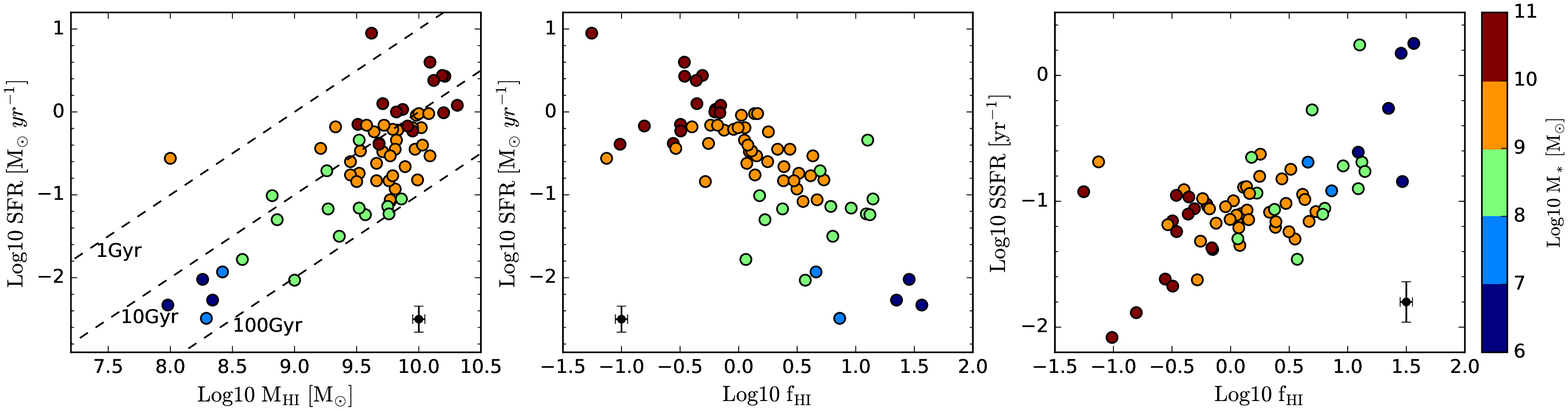}
	\caption{$M_{\hi}$-SFR (left), $f_{\hi}$-SFR (middle), and $f_{\hi}$-SSFR (right) planes for the \hi-detected sample. The dashed lines in the left panel are the timescales for depletion of the \hi\ gas reservoir, assuming a constant SFR at the current level, i.e., the reciprocal of the SFE. The galaxies in planes are binned with the stellar mass $M_*$. The mean errors are marked in each panel.}
	\label{fig_HI_SFR_GM}
\end{figure*}

\begin{figure*}
	\centering
	\includegraphics[width=1.1\hsize]{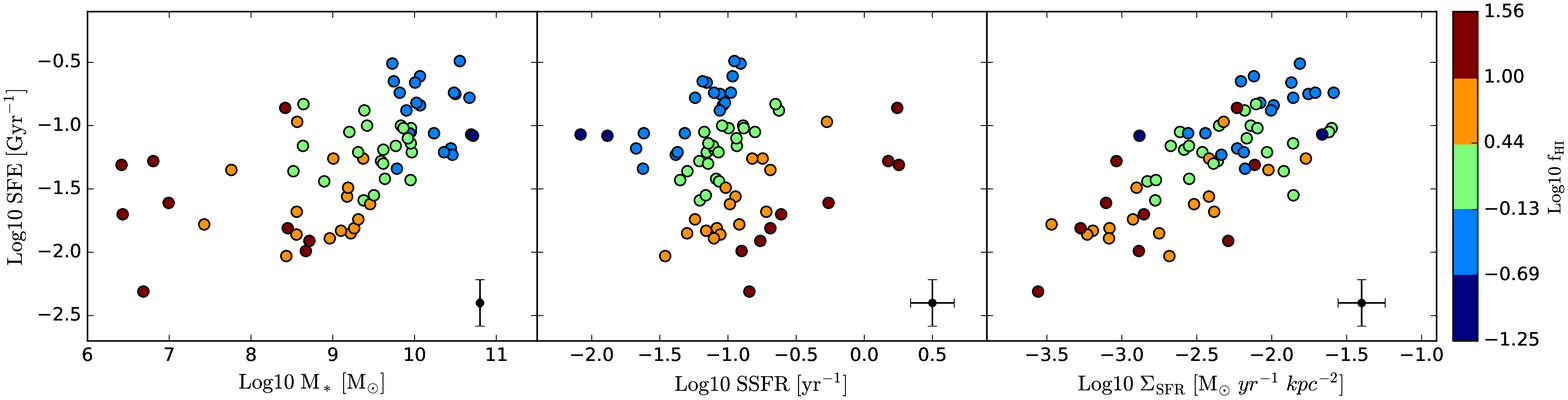}
	\caption{Relation between SFE and galaxy mass $M_*$ (left), SSFR (middle), and SFR (right). Color indicates \hi\ gas fraction $f_{\hi}$.}
	\label{fig_SFE_SFR_GM}
\end{figure*}

\subsection{Gas scaling relations of the \hi-detected galaxies}

To demonstrate the role of \hi\ gas in galaxy evolution, we further investigate the correlations  between \hi\ and physical properties of galaxies.

\subsubsection{The {\rm SFR} and {\rm SSFR} in \hi-detected Galaxies}

Figure \ref{fig_HI_SFR_GM} shows how SFR and SSFR vary with \hi\ mass ($M_{\hi}$) and \hi\ fraction ($f_{\hi} \equiv M_{\hi} / M_*$) along with the $M_*$ bins. We confirm the previous findings \citep[e.g.,][]{Huang2012} that SFRs generally increase with $M_{\hi}$ but decrease with $f_{\hi}$ and SSFRs increase with $f_{\hi}$, and these trends still exist in a given $M_*$ bins. The star formation efficiency (${\rm SFE} \equiv {\rm SFR} / M_{\hi}$) mainly varies in the range of ${\rm log_{10} SFE} = 0.5-1.5\ {\rm Gyr^{-1}}$. Besides that, we also find the scaling relations of $M_{\hi}$ increasing and $f_{\hi}$ decreasing with $M_*$.

However, there are two extreme outliers (IC 676 and IC 698) deviating from the $M_{\hi}$-SFR and $f_{\hi}$-SSFR relations in Figure \ref{fig_HI_SFR_GM}. Both galaxies have much lower $f_{\hi}$ or higher SFE than the other counterparts with similar SFR and $M_*$. IC 676 is a S0 barred galaxy with double star-forming nuclei. It is still unclear if stellar bar or/and galaxy interaction promotes the consumption of \hi\ gas and the formation of the second nucleus. IC 698 is likely a Sa galaxy with obvious irregular spiral structures in the outer disk, and is likely perturbed by galaxy interaction, which may also drive the high SFE of this galaxy.

SFE indicates the efficiency that galaxies convert the \hi gas to molecular gas and then stars.
Figure \ref{fig_SFE_SFR_GM} presents the variation of \hi-based SFE as a function of $M_*$, SSFR and $\Sigma_{SFR}$, respectively. In general, SFE increases slightly with $M_*$, and galaxies with high \hi\ gas fraction have low SFE, which may indicate a low-efficiency SF law or low-efficient \hi-to-${\rm H_2}$ conversion in these galaxies \citep{Huang2012}. There are large scatter in the SSFR-SFE relation, but SFE increases with SSFR at fixed $f_{\hi}$ bins. This may be caused by the positive correlation between SFR and SFE. As shown in the right panel of Figure \ref{fig_SFE_SFR_GM}, SFE increases with the SFR surface density $\Sigma_{\rm SFR}$, suggesting that the efficiency of \hi-to-${\rm H_2}$ conversion is high at the region with high $\Sigma_{\rm SFR}$.

\subsubsection{The relations between \hi\ and galaxy morphology}

The morphologies and kinematics of galaxies are able to provide the clues as to the process of gas accretion, and also may determine the gas properties and efficiency of converting \hi\ gas to ${\rm H_2}$ and stars \citep[e.g.,][]{Blitz2006, Powell2013}.
We derive the concentration ($C$), asymmetry ($A$) and smoothness ($S$) indices from our $R$-band and continuum-subtracted ${\rm H\alpha}$ images.

$C$ characterizes the compactness of the light distribution in a galaxy, and is defined to be the ratio of two radii, each of which contains fixed fractions of the total flux in a galaxy \citep[e.g.,][]{Kent1985,Bershady2000}. Here, we use the index $C_{42}$ defined by \citet{Kent1985}:
\begin{equation}
C_{42}=5{\rm log_{10}}(r_{80}/r_{20}),
\end{equation}	
where $r_{20}$ and $r_{80}$ are the radii which contain 20\% and 80\% of the total luminosity, respectively.

$A$ describes the rotational symmetry of a galaxy system, and is calculated by comparing the galaxy images before and after rotated with the usual rotation angle $\phi = 180\deg$ \citep{Abraham1996, Conselice2000b}:
\begin{equation}
A=\frac{\sum{\left| I_{180\degr}-I_0 \right|}}{2\sum{\left| I_0 \right|}},
\end{equation}
where $I_0$ and $I_{180\degr}$ are the intensity distribution in the original image and rotated image with rotation angle of 180$\degr$, respectively.

$S$ measures the patchiness of the light distribution in a galaxy, and is defined as the ratio of the amount of light contained in high-frequency structures to its integrated light in the galaxy \citep{Conselice2003}:
\begin{equation}
	S=10\sum{\frac{(I_0-I_{\sigma})-B}{I_0}},
\end{equation}
where $I_0$ is the flux distribution of the galaxy, $I_{\sigma}$ is also the flux distribution, but smoothed with a filter with $\sigma$ (here $\sigma$ is 0.3 times the Petrosian radius of the galaxy), and $B$ is the background value.

The $\it CAS$ parameters correlate with star formation and major merging activity, and can be used to understand the evolutionary history of a galaxy \citep{Conselice2003}. Here we use the $\it CAS$ indices from $R$-band images to characterize the stellar distribution of galaxies, and use the indices from ${\rm H\alpha}$ images to characterize the star formation distribution.

\begin{figure*}
	\centering
	\includegraphics[width=\hsize]{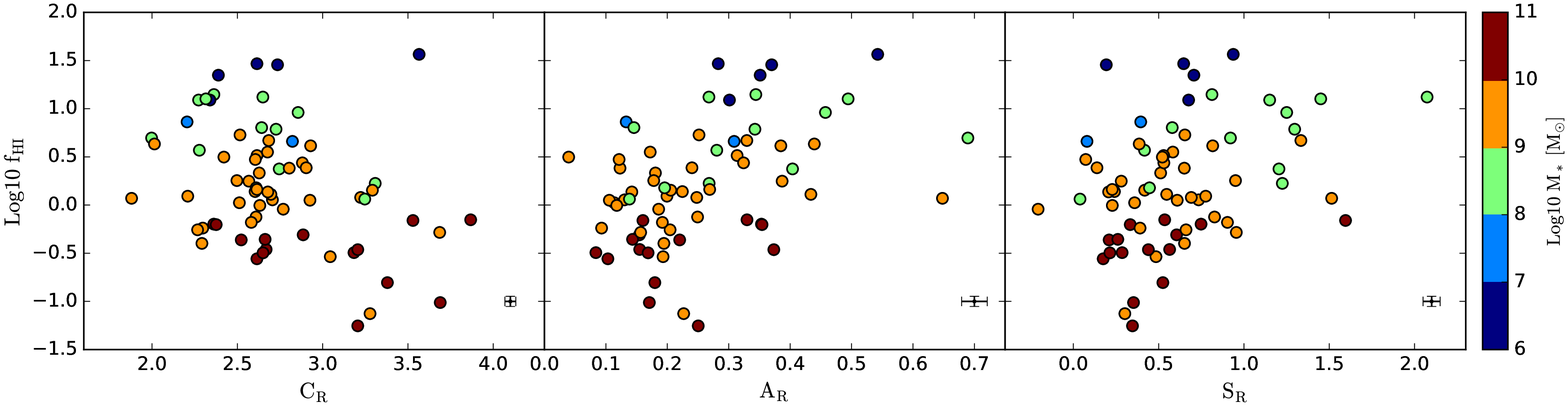}
	\includegraphics[width=\hsize]{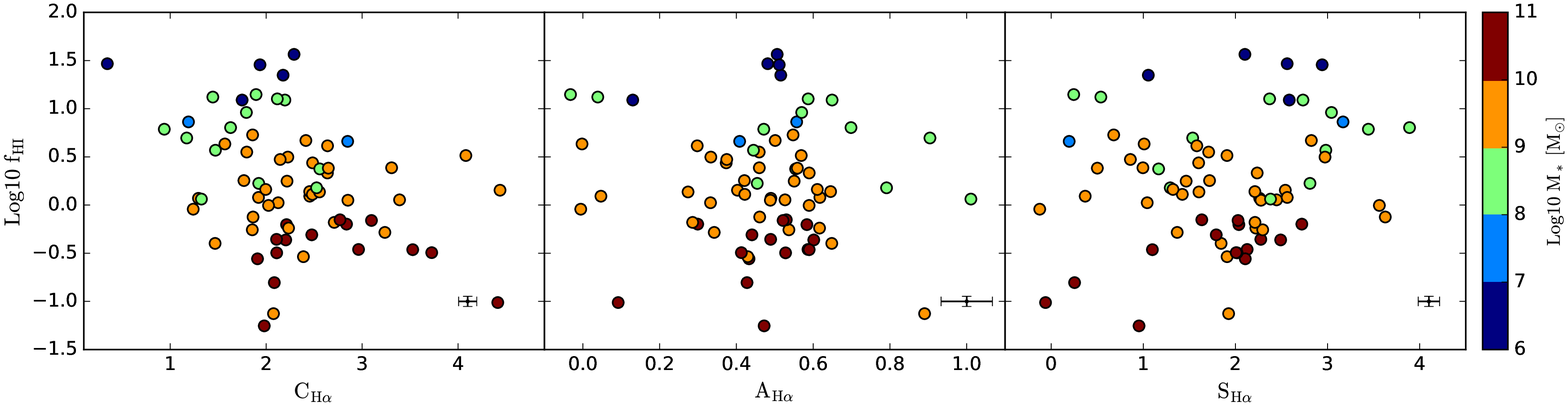}
	\caption{Relationship between \hi\ mass fraction and the morphological parameters for galaxies. The concentration ($\it C$), asymmetry ($\it A$) and smoothness ($\it S$) indices are used and measured from the observed $R$-band and continuum-subtracted ${\rm H\alpha}$ images, and characterize the distributions of stellar continuum (top panels) and star formation (bottom panels). Similar to Figure~\ref{fig_HI_SFR_GM}, the galaxies in each plane are binned with the stellar mass $M_*$.}
	\label{fig10}
\end{figure*}
\begin{figure*}
	\centering
	\includegraphics[width=\hsize]{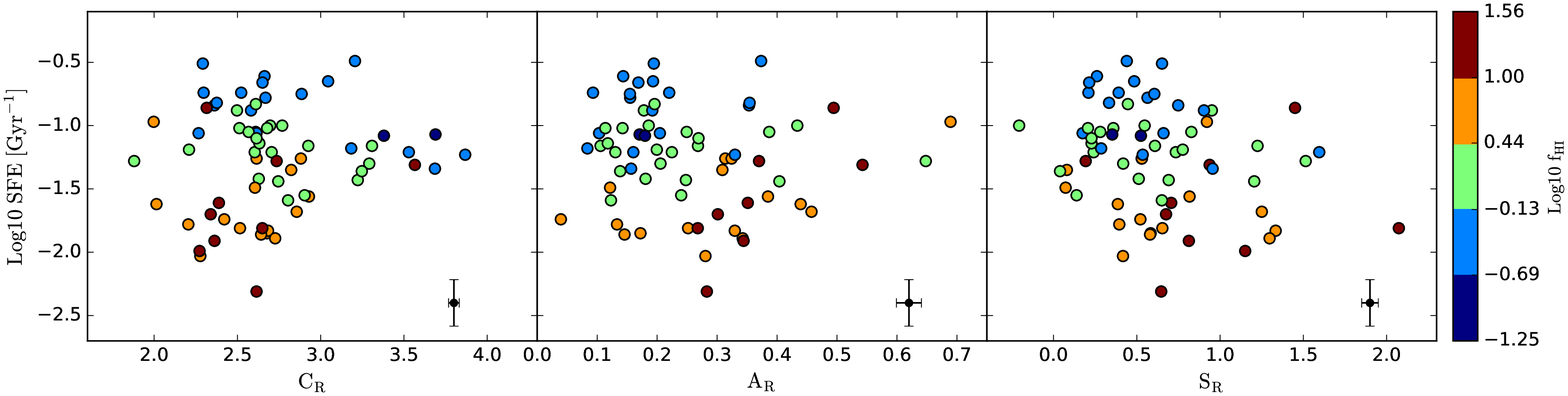}
	\includegraphics[width=\hsize]{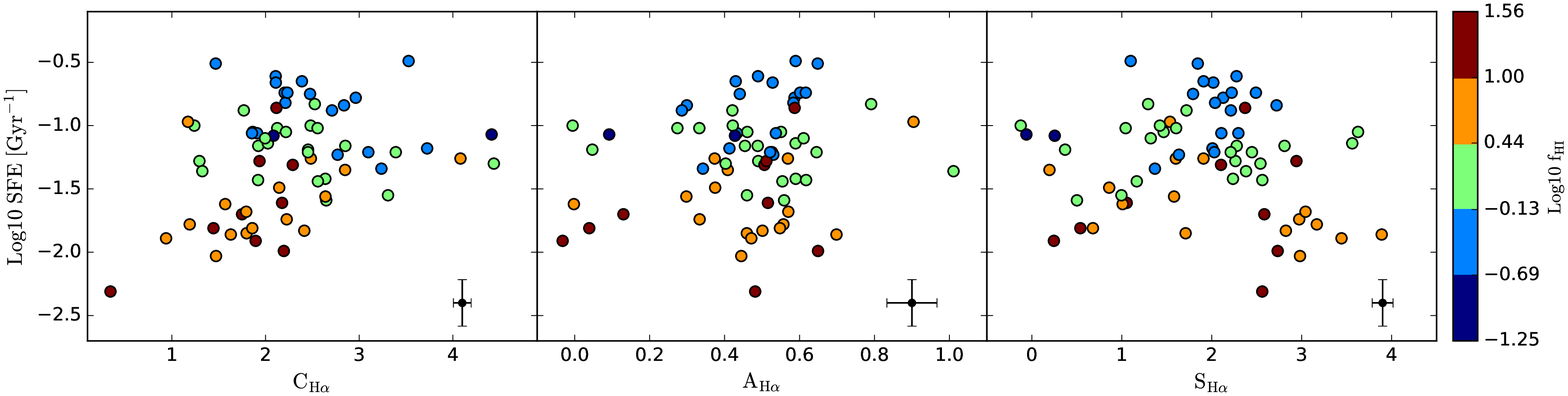}
	\caption{Relationship between \hi-based SFE and the morphological parameters for galaxies . The galaxies in each plane are binned with \hi\ mass fraction $f_{\hi}$.}
	\label{fig_SFE_CAS}
\end{figure*}

Figure \ref{fig10} presents the correlations of \hi\ fraction $f_{\hi}$ and the $\it CAS$ indices for our \hi-detected galaxies. It can be seen that, in general, galaxies with lower concentration or higher asymmetry have higher gas fraction, although the dependence becomes very weak at a given stellar mass. The mild trend of \hi\ fraction with the $\it CAS$ is likely an indication of external processes or mechanisms that control the gas accretion and supply \citep{Schiminovich2010}.
In the bottom three panels of Figure \ref{fig10}, we find little or no obvious trends between $f_{\hi}$ and $\it CAS$ indices for star formation distribution in galaxies except $f_{\hi}$ vs. $\it C_{\rm H\alpha}$, but which is likely driven by the stellar mass. It may suggest that there is little correlation between the spacial distribution of star formation activity and gas fraction due to the smooth process of gas accretion \citep{Wang2011}.

In order to explore the effect of galaxy morphology on SFE, we plot \hi-based SFE on a function of $\it CAS$ indices in Figure \ref{fig_SFE_CAS}. In each panel, the galaxies are binned with the gas fraction. However, in most of the panels, we find little trends between SFE and the morphology index of stellar mass and star formation. Only weak correlation exists between SFE and concentration $C$, especially for $C_{\rm H{\alpha}}$. For the galaxies with $C<3.0$, SFE appears to increase with $C$. While for the objects with $C>3.0$, the trend is seemingly reversed. The discrepancy between the galaxies with high and low concentration is similar to the display between \hi\ depletion times and stellar surface density found by \citet{Jaskot2015}, and is likely due to the quenching of star formation in more massive galaxies.

\section{Discussion}
\label{sec:discuss}

Atomic gas plays an important role in the evolution of galaxies by providing the fuel for star formation activity. The study of \hi is essential to gaining a better understanding of the main physical processes in galaxy evolution. In the current work, we have derived the properties of \hi-detected galaxies in a ALFALFA sky region, and investigated relations between \hi\ gas content (mass $M_{\hi}$, fraction $f_{\hi}$ and SFE) and global galaxy properties such as star formation activity and morphology.

There have been many studies into the question of how the \hi properties of galaxies vary with star formation, galaxy structure, and other physical properties \citep[e.g.,][]{Roberts1994, Haynes2011, Toribio2011a, Huang2012}. As we found in Figures \ref{fig_dist} and \ref{fig_hist}, \hi-detected galaxies are dominated by blue star-forming galaxies, and are strongly biased against the red sequence in the color-magnitude diagram. Many studies have pointed out that there are significant differences between the distributions of physical properties for \hi-detected and -undetected galaxies, such as stellar mass surface density, star formation activity, and morphological index \citep{West2010,Haynes2011}.
Similarly, \citet{Huang2012} compared the \hi\ and optically selected samples over the same restricted volume, and found that the \hi-selected population is relatively less evolved and has overall higher SFR, higher SSFR, and bluer colors at a given stellar mass, but lower SFE and extinction.

These results are further reinforced by the distributions of our samples in the SFR and SSFR versus stellar mass diagrams. As Figure ~\ref{fig_SFR_GM} shows, most of \hi-detected galaxies are in good agreement with the star-forming main sequence. In addition, the SSFRs of the our sample show a much shallower decline with increasing stellar mass than the normal non-star-forming galaxies. In agreement with our results, \citet{Bauer2013} found that the dependence of SSFR on stellar mass is steeper for the full sample of galaxies than it is for just the star-forming population. \citet{Gavazzi2013b} found a similar trend in the SSFR-mass diagrams, and identified a four-step sequence of the galaxy quenching process including \hi-rich late-type galaxies, \hi-poor late-type galaxies, \hi-poor galaxies with little star formation, but post-starburst signature, and early-type galaxies.

Recent investigations into the evolutionary properties of galaxies have tried to find possible mechanisms for transforming galaxies from blue actively star-forming systems into red passive systems \citep[e.g.,][]{Bothwell2009, Peng2015, Davies2016b}. Because \hi\ gas is the initial reservoir where molecular gas and stars are formed subsequently, the star formation in the main Sequence galaxies is supposed to be the produce of relatively stable processes, such as gas inflow and accretion \citep{Davies2016a}, and the low SFRs of the red sequence result from a lack of available gas supply \citep{Bothwell2009}.
\citet{Saintonge2016} indicated that if there are no extended \hi\ envelopes or other sources of accretion to replenish the gas reservoirs, galaxies in the main sequence will cease actively forming stars and migrate to the red cloud along with the simultaneous processes in the transformation, such as reduction of the gas fractions, ageing of stellar population and growth of central bulges.

Consistent with these results, Figure \ref{fig_HI_SFR_GM} shows that SFRs generally increase with \hi\ mass, SSFRs increase with \hi\ fraction even for the galaxies in given stellar masses, and further confirms the dependence of star formation in galaxies on \hi\ gas. The \hi-based SFE also shows an increasing function of stellar mass $M_*$ and $\Sigma_{\rm SFR}$ (Figure \ref{fig_SFE_SFR_GM}). Although \citet{Huang2012} found that SFE remains relatively constant in the high $M_*$ range, we did not find this trend limiting to our sample.

In addition, the $f_{\hi}$ and SFE of \hi-selected galaxies also show correlations with the distribution of stellar mass and star formation in galaxies. As illustrated in Figures \ref{fig10} and \ref{fig_SFE_CAS}, $f_{\hi}$ mildly increases with the asymmetry and decreases with the concentration of galaxies, and SFE mildly increases with the concentration $C_{\rm H\alpha}$ at the lower end. This may suggest that the \hi\ gas supply and effect are more correlated with external processes or extended disks of galaxies \citep{Wang2011}. Extended galaxies generally have low concentration for the distributions of their stellar mass and star formation, and external processes accrete their ISM to the outer disk which is dominated by \hi\ gas. As a result, the galaxies with lower concentration will have higher gas fraction. Furthermore, SFE decreases with radius across the outer disks of spirals \citep{Bigiel2010b}, thus lower SFE is expected in the galaxies with lower concentration.


While the atomic gas shows a strong correlation with star formation on global scales of galaxies, there seems to be little or no relations between them on smaller scales because star formation is more directly related to molecular gas than atomic gas. As we showed in Figures \ref{fig10} and \ref{fig_SFE_CAS}, $f_{\hi}$ and SFE have relatively weak or no connections with the distributions of star formation in galaxies except for the concentration,
and the cause might be that the gas richness of galaxies is more closely correlated with the past-averaged star formation in the last Gyr than with the current star formation \citep{Catinella2010, Kannappan2013}.
In addition, more detailed studies on sub-kpc scales show the lack of correlation between surface densities of SFR and \hi\ \citep[e.g.,][]{Kennicutt2007, Bigiel2008}. \citet{Wang2017b} indicated that the apparently different behaviors of \hi-SFR relations on different scales may be caused by side effects, i.e., \hi\ and SFR both correlate with other physical parameters of galaxies, such as galaxy size and stellar mass, but the physical link between them is intrinsically weak.

Owing to the fact that stars are formed in molecular regions on parsec scales, the above results were somewhat expected. However, \hi\ gas still plays an important role in governing the star formation activity \citep{Fumagalli2008, Bigiel2010b}. Besides regulating the formation of giant molecular clouds \citep{Blitz2004}, \hi\ is found to correlate with star formation in the outer disks of spirals and in dwarf galaxies where the ISM is dominated by neutral gas \citep{Bigiel2010a, Wang2011}.
In the investigations of the star formation Schmidt law, \citet{Liu2015} found the total gas density ($\rm H_2 + \hi$) shows a steeper function with SFR density than only molecular gas density, especially at the low density regime of molecular gas, indicating that atomic gas may have indirect effect on SF.

While many observational studies have been made, current results for the relation of \hi\ and star formation are still limited by the relatively small number of statistics and lack of deeper data. The star formation tracers used in current works, such as UV, IR, and H$\alpha$ observations, usually have not enough detection limiting, spacial resolution, or sky coverage. Thanks to the recent and upcoming deep surveys, such as the South Galactic Cap {\it u}-band Sky Survey \citep[SCUSS;][]{ZhouX2016,Zou2015}, which provide opportunities to measure SFRs of unprecedented large samples with optical {\it u}-band data \citep{Zhou2017}. Hence, analyses of larger samples of galaxies based on SCUSS and \hi\ data will be the subject of our future work.

\section{Summary}
\label{sec:summary}

We have investigated the correlations of \hi\ gas and physical properties of galaxies with the aim of exploring the role of \hi\ in governing galaxy evolution. We select all galaxies detected by ALFALFA in a 15 deg$^2$ sky region as the \hi-detected sample, and observe the H$\alpha$ narrow-band imaging to derive their star formation properties. We derive the physical properties of the galaxies and analyze the dependence of star formation and structure on \hi\ gas.
Our main results can be summarized as follows:

\begin{enumerate}
\item{} We have obtained $M_*$, SFR, SSFR, $\Sigma_{\rm SFR}$, $\Sigma_{M_*}$ and concentration index of \hi-detected galaxies. In general, our \hi-detected galaxies have low $\Sigma_{M_*}$ and active star formation, consistent with previous findings. Most of the galaxies are in good agreement with the star-forming main sequence in the SFR-$M_*$ plane, and show shallower dependence of SSFR on stellar mass than non-star-forming samples.

\item{} The dependence of star formation in galaxies on \hi\ gas is confirmed at least on global scale: SFRs generally increase with \hi\ mass but decrease with \hi\ fraction, SSFRs increase with \hi\ fraction even for the galaxies in given stellar mass bins. \hi-based SFE is found to increase with the stellar mass and SFR surface density.

\item{} $f_{\hi}$ and SFE of \hi-selected galaxies show correlations with the distribution of stellar mass and star formation in galaxies. $f_{\hi}$ mildly increases with the asymmetry and decreases with the concentration of galaxies, and SFE mildly increases with the concentration $C_{\rm H\alpha}$ at the lower end, suggesting that the \hi\ gas supply and effect is likely more correlated with external processes or extended disks of galaxies.

\item{} There are relatively weak or no trends found between $f_{\hi}$, SFE and other morphological indices except the concentration for the distributions of stellar mass and star formation in galaxies, which may be caused by the weak physical link between \hi\ and star formation in small scale and the smooth accretion process of \hi\ gas.
\end{enumerate}

\acknowledgements
\label{sec:acknow}
We thank the anonymous referee for numerous valuable suggestions.
We gratefully thank Q. Yuan, J. Wang, and C. Lee for useful discussions.
This work was supported by the Chinese National Natural Science Foundation grands Nos. 11433005, 11733006, 11673027, 11373035, and 11603034, by National Key R\&D Program of China (grant No. 2017YFA0402704) and by the National Basic Research Program of China (973 Program), No. 2014CB845704, 2013CB834902, 2014CB845702, and 2015CB857004.

We acknowledge the support of the kind staff of Lijiang 2.4 m and Xinglong 2.16 m telescopes. Funding for the Lijiang 2.4 m telescope is provided by the Chinese Academy of Sciences and the People's Government of Yunnan Province.
The authors would like to acknowledge the work of the entire ALFALFA collaboration team in observing, flagging, and extracting the catalog of galaxies used in this work. Funding for the SDSS and SDSS-II has been provided by the Alfred P. Sloan Foundation, the Participating Institutions, the National Science Foundation, the U.S. Department of Energy, the National Aeronautics and Space Administration, the Japanese Monbukagakusho, the Max Planck Society, and the Higher Education Funding Council for England. The SDSS website is http://www.sdss.org/. The SDSS MPA-JHU catalog was produced by a collaboration of researchers (currently or formerly) from the MPA and the JHU. The team is made up of Stephane Charlot, Guinevere Kauffmann and Simon White (MPA), Tim Heckman (JHU), Christy Tremonti (University of Arizona - formerly JHU), and Jarle Brinchmann (Centro de Astrof\'isica da Universidade do Porto - formerly MPA). This publication also makes use of data products from the $\it WISE$, which is a joint project of the University of California, Los Angeles, and the Jet Propulsion Laboratory/California Institute of Technology, funded by the National Aeronautics and Space Administration.

\bibliographystyle{apj}
\bibliography{}

\appendix
\section{The Atlas}
An Atlas of the 70 galaxies in our sample is given in this Appendix. Table \ref{table:obsevations} provides the information of the galaxies and the fluxes of \hi\, $\rm H\alpha$ and WISE IR emission. Table \ref{table:properties} gives the physical properties and morphological parameters for galaxies. Figure \ref{fig_imageshow} displayed the images in broadband $R$, continuum-subtracted narrowband H$\alpha$, WISE IR 3.4$\mu$m and 12$\mu$m, along with the \hi\ Spectrum from ALFALFA.

\renewcommand\thetable{\thesection.\arabic{table}}
\startlongtable
\begin{deluxetable}{rrrrrllll}
\setcounter{table}{0}
\tablecaption{observational data of the target galaxies.\label{table:obsevations}}
\tablehead{	\colhead{AGC} & \colhead{R.A.} & \colhead{Decl.} & \colhead{$cz$} & \colhead{Distance} & \colhead{$S_{21}$} & \colhead{${\rm log}_{10} F_{3.4\mu m}$}  & \colhead{${\rm log}_{10} F_{12\mu m}$} & \colhead{$log_{10} F_{H\alpha}$}  \\
	\colhead{} & \colhead{(deg)} & \colhead{(deg)} & \colhead{(km s$^{-1}$)} & \colhead{(Mpc)} & \colhead{(km s$^{-1}$)} & \colhead{(erg s$^{-1}$ cm$^{-2}$)} & \colhead{(erg s$^{-1}$ cm$^{-2}$)} & \colhead{(erg s$^{-1}$ cm$^{-2}$)} \\
	\colhead{(1)} & \colhead{(2)} & \colhead{(3)} & \colhead{(4)} & \colhead{(5)} & \colhead{(6)} & \colhead{(7)} & \colhead{(8)} & \colhead{(9)}}
\startdata
  200803 & 165.04793 & 10.36444 &  11018 & 162.4 $\pm$ 10.9 & 1.18 $\pm$  0.18 & -11.88 $\pm$  0.01 & -11.79 $\pm$ 0.01 & -13.40 $\pm$ 0.03\\
  200805 & 165.09541 & 9.84944  &  9296  & 137.8 $\pm$ 9.3 & 1.31 $\pm$  0.15 & -12.50 $\pm$  0.01 & -12.84 $\pm$ 0.12 & -13.54 $\pm$ 0.03\\
  200845 & 165.48582 & 10.29500 &  10417 & 153.8 $\pm$ 10.4 & 1.89 $\pm$  0.24 & -11.93 $\pm$  0.01 & -11.97 $\pm$ 0.01 & -12.99 $\pm$ 0.03\\
  200847 & 165.51083 & 10.34417 &  10721 & 158.2 $\pm$ 10.8 & 0.78 $\pm$  0.11 & -12.29 $\pm$  0.01 & -12.44 $\pm$ 0.05 & -13.26 $\pm$ 0.04\\
  200855 & 165.59958 & 10.34333 &  11327 & 166.8 $\pm$ 11.2 & 1.41 $\pm$  0.20 & -12.29 $\pm$  0.01 & -12.32 $\pm$ 0.04 & -13.66 $\pm$ 0.03\\
  200858 & 165.62459 & 9.98861  &  10682 & 157.6 $\pm$ 10.6 & 0.74 $\pm$  0.14 & -12.47 $\pm$  0.01 & -12.54 $\pm$ 0.07 & -13.59 $\pm$ 0.02\\
  210063 & 166.95834 & 10.81361 &  9089  & 134.9 $\pm$ 9.1  & 0.50 $\pm$  0.12 & -12.03 $\pm$  0.01 & -12.07 $\pm$ 0.03 & -13.51 $\pm$ 0.02\\
  210072 & 167.08084 & 10.0525  &  12255 & 180.1 $\pm$ 12.1 & 2.14 $\pm$  0.21 & -11.44 $\pm$  0.01 & -11.56 $\pm$ 0.01 & -13.43 $\pm$ 0.03\\
  210073 & 167.07916 & 9.95694  &  12067 & 177.4 $\pm$ 12.0 & 2.11 $\pm$  0.20 & -11.58 $\pm$  0.01 & -11.45 $\pm$ 0.01 & -13.20 $\pm$ 0.01\\
  210082 & 167.35042 & 10.83667 &  1555  & 17.5  $\pm$ 1.4  & 2.72 $\pm$  0.10 & -12.31 $\pm$  0.01 & -12.68 $\pm$ 0.08 & -13.32 $\pm$ 0.02\\
  210084 & 167.35791 & 9.76972  &  7862  & 117.4 $\pm$ 7.9  & 1.57 $\pm$  0.15 & -12.00 $\pm$  0.01 & -12.25 $\pm$ 0.02 & -13.87 $\pm$ 0.04\\
  210111 & 167.60583 & 10.12278 &  1320  & 17.5  $\pm$ 1.2  & 3.02 $\pm$  0.13 & -12.70 $\pm$  0.01 & -12.55 $\pm$ 0.06 & -13.88 $\pm$ 0.03\\
  210148 & 167.91417 & 9.69444  &  13847 & 202.9 $\pm$ 13.6 & 0.69 $\pm$  0.10 & -12.18 $\pm$  0.01 & -12.23 $\pm$ 0.03 & -13.79 $\pm$ 0.10\\
  210171 & 168.36581 & 10.48472 &  8885  & 131.9 $\pm$ 8.8  & 1.16 $\pm$  0.13 & -11.55 $\pm$  0.01 & -12.04 $\pm$ 0.02 & -13.39 $\pm$ 0.02\\
  210180 & 168.68459 & 10.22833 &  13958 & 204.4 $\pm$ 13.7 & 1.34 $\pm$  0.18 & -11.71 $\pm$  0.01 & -11.64 $\pm$ 0.01 & -13.47 $\pm$ 0.02\\
  210354 & 172.06084 & 9.08278  &  6254  & 94.4  $\pm$ 6.3  & 4.58 $\pm$  0.17 & -11.51 $\pm$  0.01 & -11.40 $\pm$ 0.01 & -12.70 $\pm$ 0.00\\
  210368 & 172.25792 & 9.10083  &  6287  & 94.9  $\pm$ 6.4  & 1.55 $\pm$  0.11 & -12.37 $\pm$  0.01 & -12.64 $\pm$ 0.08 & -13.76 $\pm$ 0.04\\
  211086 & 167.12582 & 10.80806 &  10978 & 161.8 $\pm$ 10.9 & 1.07 $\pm$  0.17 & -12.15 $\pm$  0.01 & -12.16 $\pm$ 0.03 & -13.79 $\pm$ 0.03\\
  212984 & 165.99001 & 10.10806 &  9907  & 146.6 $\pm$ 9.8  & 1.21 $\pm$  0.14 & -12.00 $\pm$  0.01 & -12.02 $\pm$ 0.02 & -14.24 $\pm$ 0.10\\
  212987 & 166.30249 & 10.44778 &  9220  & 136.7 $\pm$ 9.3  & 1.34 $\pm$  0.12 & -12.60 $\pm$  0.01 & -12.90 $\pm$ 0.12 & -13.46 $\pm$ 0.05\\
  212988 & 166.53793 & 10.03222 &  5776  & 87.5  $\pm$ 6.0  & 0.40 $\pm$  0.11 & -12.63 $\pm$  0.01 & -12.98 $\pm$ 0.14 & -14.29 $\pm$ 0.06\\
  212989 & 166.53625 & 10.14528 &  14349 & 210.0 $\pm$ 14.1 & 0.64 $\pm$  0.13 & -12.15 $\pm$  0.01 & -12.32 $\pm$ 0.04 & -13.77 $\pm$ 0.02\\
  212990 & 166.79501 & 10.26639 &  9005  & 133.7 $\pm$ 9.0  & 2.33 $\pm$  0.11 & -12.43 $\pm$  0.01 & -12.49 $\pm$ 0.05 & -13.89 $\pm$ 0.08\\
  212994 & 167.10957 & 10.48944 &  13805 & 202.2 $\pm$ 13.6 & 0.53 $\pm$  0.12 & -12.08 $\pm$  0.01 & -12.24 $\pm$ 0.03 & -13.62 $\pm$ 0.01\\
  212996 & 167.46541 & 10.40083 &  8928  & 132.6 $\pm$ 8.9  & 1.55 $\pm$  0.19 & -12.33 $\pm$  0.02 & -12.38 $\pm$ 0.04 & -13.90 $\pm$ 0.02\\
  213010 & 170.49083 & 10.49444 &  4461  & 66.2  $\pm$ 4.8  & 0.96 $\pm$  0.10 & -12.57 $\pm$  0.01 & -13.05 $\pm$ 4.34 & -15.71 $\pm$ 3.10\\
  213058 & 165.63792 & 9.47194  &  8241  & 122.8 $\pm$ 8.2  & 0.92 $\pm$  0.19 & -11.70 $\pm$  0.01 & -11.96 $\pm$ 0.01 & -13.54 $\pm$ 0.01\\
  213062 & 166.77167 & 9.80222  &  6254  & 94.4  $\pm$ 6.3  & 0.88 $\pm$  0.06 & -12.46 $\pm$  0.01 & -12.59 $\pm$ 0.07 & -13.56 $\pm$ 0.05\\
  213064 & 167.72626 & 9.62278  &  1604  & 17.5  $\pm$ 1.4  & 3.65 $\pm$  0.10 & -12.01 $\pm$  0.01 & -12.31 $\pm$ 0.03 & -12.68 $\pm$ 0.01\\
  213069 & 168.20749 & 9.52278  &  6250  & 94.3  $\pm$ 6.3  & 1.10 $\pm$  0.11 & -12.76 $\pm$  0.01 & -12.85 $\pm$ 0.36 & -14.62 $\pm$ 0.37\\
  213072 & 169.71918 & 9.62889  &  9278  & 137.6 $\pm$ 9.2  & 0.86 $\pm$  0.05 & -11.96 $\pm$  0.01 & -11.84 $\pm$ 0.01 & -13.46 $\pm$ 0.01\\
  213074 & 169.86583 & 9.59917  &  990   & 13.7  $\pm$ 0.9  & 2.14 $\pm$  0.05 & -13.00 $\pm$  0.02 & -13.13 $\pm$ 1.00 & -13.40 $\pm$ 0.01\\
  213076 & 170.32251 & 9.15194  &  8981  & 133.3 $\pm$ 8.9  & 1.55 $\pm$  0.13 & -12.39 $\pm$  0.01 & -12.59 $\pm$ 0.06 & -14.15 $\pm$ 0.09\\
  213083 & 170.86749 & 9.08750  &  9456  & 140.1 $\pm$ 9.4  & 0.61 $\pm$  0.12 & -12.43 $\pm$  0.01 & -12.64 $\pm$ 0.07 & -14.39 $\pm$ 0.06\\
  213085 & 171.63333 & 9.91945  &  9391  & 139.2 $\pm$ 9.3  & 1.22 $\pm$  0.11 & -12.74 $\pm$  0.01 & -12.84 $\pm$ 0.25 & -14.20 $\pm$ 0.25\\
  213087 & 172.05290 & 9.80417  &  9321  & 138.2 $\pm$ 9.4  & 1.31 $\pm$  0.15 & -12.15 $\pm$  0.01 & -12.39 $\pm$ 0.04 & -13.90 $\pm$ 0.04\\
  213090 & 172.36917 & 10.62361 &  5855  & 88.6  $\pm$ 5.9  & 0.98 $\pm$  0.06 & -12.70 $\pm$  0.01 & -12.79 $\pm$ 0.11 & -13.43 $\pm$ 0.01\\
  213092 & 172.41000 & 9.96444  &  5887  & 89.1  $\pm$ 6.2  & 1.51 $\pm$  0.13 & -12.14 $\pm$  0.01 & -12.12 $\pm$ 0.03 & -13.63 $\pm$ 0.01\\
  213198 & 168.51875 & 10.98972 &  12357 & 181.5 $\pm$ 12.2 & 0.61 $\pm$  0.09 & -12.10 $\pm$  0.01 & -12.34 $\pm$ 0.04 & -13.40 $\pm$ 0.02\\
  213573 & 165.16417 & 9.22806  &  10723 & 158.2 $\pm$ 10.6 & 0.97 $\pm$  0.14 & -12.48 $\pm$  0.01 & -12.67 $\pm$ 0.08 & -13.75 $\pm$ 0.03\\
  213574 & 165.16917 & 9.04778  &  10651 & 157.2 $\pm$ 10.6 & 0.28 $\pm$  0.06 & -12.14 $\pm$  0.01 & -12.38 $\pm$ 0.06 & -14.16 $\pm$ 0.02\\
  213583 & 165.39708 & 9.38944  &  11826 & 174.0 $\pm$ 11.7 & 0.73 $\pm$  0.09 & -12.09 $\pm$  0.01 & -12.05 $\pm$ 0.02 & -13.32 $\pm$ 0.03\\
  213586 & 165.97125 & 9.13944  &  12171 & 178.9 $\pm$ 12.0 & 0.63 $\pm$  0.12 & -11.41 $\pm$  0.00 & -12.73 $\pm$ 0.07 & -13.85 $\pm$ 0.01\\
  213588 & 166.53084 & 9.63806  &  12358 & 181.6 $\pm$ 12.2 & 0.79 $\pm$  0.11 & -12.78 $\pm$  0.01 & -12.88 $\pm$ 0.25 & -13.45 $\pm$ 0.09\\
  213590 & 167.12709 & 9.92556  &  12027 & 176.8 $\pm$ 11.9 & 1.67 $\pm$  0.17 & -12.50 $\pm$  0.01 & -12.52 $\pm$ 0.05 & -13.36 $\pm$ 0.01\\
  213596 & 167.57959 & 9.50917  &  13260 & 194.5 $\pm$ 13.1 & 1.00 $\pm$  0.15 & -12.14 $\pm$  0.01 & -12.16 $\pm$ 0.02 & -14.57 $\pm$ 0.22\\
  213603 & 167.99208 & 9.46028  &  14080 & 206.2 $\pm$ 13.8 & 0.34 $\pm$  0.06 & -12.67 $\pm$  0.01 & -13.06 $\pm$ 0.20 & -14.15 $\pm$ 0.02\\
  213609 & 168.20876 & 9.73389  &  14033 & 205.5 $\pm$ 13.8 & 1.01 $\pm$  0.15 & -12.27 $\pm$  0.01 & -12.40 $\pm$ 0.05 & -13.73 $\pm$ 0.01\\
  213611 & 168.47542 & 9.64583  &  12370 & 181.8 $\pm$ 12.2 & 1.14 $\pm$  0.10 & -11.65 $\pm$  0.00 & -12.19 $\pm$ 0.02 & -13.38 $\pm$ 0.02\\
  215236 & 166.10249 & 10.26389 &  9314  & 138.1 $\pm$ 9.3  & 1.29 $\pm$  0.16 & -12.99 $\pm$  0.02 & -13.03 $\pm$ 0.21 & -13.98 $\pm$ 0.04\\
  215237 & 166.35249 & 10.67500 &  9233  & 136.9 $\pm$ 9.2  & 0.84 $\pm$  0.14 & -13.18 $\pm$  0.03 & -13.03 $\pm$ 0.17 & -13.78 $\pm$ 0.08\\
  215239 & 167.27249 & 10.16722 &  11916 & 175.3 $\pm$ 11.7 & 1.01 $\pm$  0.11 & -13.16 $\pm$  0.03 & -13.11 $\pm$ 0.86 & -13.87 $\pm$ 0.04\\
  215240 & 168.47542 & 9.93583  &  1610  & 17.5  $\pm$ 1.4  & 0.46 $\pm$  0.06 & -13.20 $\pm$  0.03 & -13.07 $\pm$ 1.00 & -15.39 $\pm$ 1.97\\
  215241 & 169.25917 & 10.14778 &  1765  & 17.5  $\pm$ 1.8  & 1.97 $\pm$  0.11 & -12.97 $\pm$  0.02 & -13.09 $\pm$ 2.10 & -15.05 $\pm$ 1.17\\
  215242 & 167.09500 & 10.08333 &  9090  & 134.9 $\pm$ 9.1  & 0.78 $\pm$  0.11 & -13.07 $\pm$  0.02 & -12.83 $\pm$ 0.24 & -14.37 $\pm$ 0.18\\
  215761 & 165.64250 & 9.48167  &  10486 & 154.8 $\pm$ 10.4 & 1.36 $\pm$  0.16 & -12.35 $\pm$  0.01 & -12.55 $\pm$ 0.05 & -13.27 $\pm$ 0.01\\
  219116 & 165.89792 & 9.29194  &  8226  & 122.6 $\pm$ 8.2  & 0.93 $\pm$  0.10 & -13.11 $\pm$  0.03 & -12.93 $\pm$ 0.54 & -13.65 $\pm$ 0.02\\
  6086   & 165.14334 & 9.95333  &  9304  & 137.9 $\pm$ 9.3  & 2.40 $\pm$  0.31 & -11.85 $\pm$  0.01 & -12.09 $\pm$ 0.02 & -13.41 $\pm$ 0.03\\
  6091   & 165.16708 & 9.87639  &  10715 & 158.1 $\pm$ 10.6 & 3.49 $\pm$  0.20 & -11.51 $\pm$  0.01 & -11.86 $\pm$ 0.01 & -13.02 $\pm$ 0.01\\
  6093   & 165.20375 & 10.73056 &  10806 & 159.4 $\pm$ 10.8 & 1.36 $\pm$  0.15 & -11.29 $\pm$  0.01 & -12.05 $\pm$ 0.02 & -12.87 $\pm$ 0.01\\
  6209   & 167.48624 & 10.72000 &  1584  & 18.1 $\pm$ 1.4  & 8.58 $\pm$  0.25 & -11.25 $\pm$  0.01 & -11.16 $\pm$ 0.00 & -11.95 $\pm$ 0.01\\
  6245   & 168.17125 & 9.06528  &  1421  & 17.5 $\pm$ 1.2  & 1.39 $\pm$  0.12 & -10.79 $\pm$  0.01 & -10.45 $\pm$ 0.00 & -12.37 $\pm$ 0.01\\
  6248   & 168.21750 & 10.19972 &  1286  & 17.5 $\pm$ 1.1  & 2.53 $\pm$  0.05 & -12.86 $\pm$  0.02 & -13.08 $\pm$ 0.19 & -13.65 $\pm$ 0.05\\
  6288   & 169.01500 & 10.17083 &  5897  & 89.3 $\pm$ 6.0  & 2.46 $\pm$  0.14 & -12.15 $\pm$  0.01 & -12.37 $\pm$ 0.04 & -13.29 $\pm$ 0.01\\
  6438   & 171.47292 & 9.98472  &  1156  & 20.4 $\pm$ 1.5  & 3.88 $\pm$  0.07 & -11.47 $\pm$  0.01 & -11.87 $\pm$ 0.01 & -13.54 $\pm$ 0.01\\
  6470   & 172.05916 & 9.14778  &  6311  & 95.2 $\pm$ 6.4  & 5.77 $\pm$  0.20 & -10.99 $\pm$  0.01 & -10.88 $\pm$ 0.00 & -12.58 $\pm$ 0.01\\
  6474   & 172.09750 & 9.41083  &  1716  & 33.1 $\pm$ 2.1  & 12.37$\pm$  0.21 & -10.76 $\pm$  0.01 & -11.43 $\pm$ 0.00 & -12.28 $\pm$ 0.01\\
  6475   & 172.14458 & 9.09805  &  6309  & 95.2 $\pm$ 6.5  & 7.40 $\pm$  0.22 & -11.16 $\pm$  0.01 & -11.43 $\pm$ 0.00 & -12.99 $\pm$ 0.01\\
  6477   & 172.15457 & 9.09972  &  6288  & 94.9 $\pm$ 6.5  & 5.63 $\pm$  0.13 & -11.55 $\pm$  0.01 & -11.57 $\pm$ 0.01 & -12.60 $\pm$ 0.01\\
  6482   & 172.26625 & 9.10639  &  6201  & 93.6 $\pm$ 6.3  & 2.03 $\pm$  0.22 & -10.69 $\pm$  0.01 & -10.40 $\pm$ 0.00 & -12.58 $\pm$ 0.01\\
\enddata
\end{deluxetable}

\renewcommand\thetable{\thesection.\arabic{table}}
\startlongtable
\begin{deluxetable*}{ccccrrrrrrr}
\setcounter{table}{1}
\tablecaption{Physical properties of the target galaxies.\label{table:properties}}
\tablehead{\colhead{AGC} & \colhead{${\rm log}_{10} M_{\hi}$} & \colhead{${\rm log}_{10} M_*$} & \colhead{${\rm log}_{10} {\rm SFR}$} & \colhead{$C_R$} & \colhead{$A_R$} & \colhead{$S_R$} & \colhead{$C_{\rm H\alpha}$} & \colhead{$A_{\rm H\alpha}$} & \colhead{$S_{\rm H\alpha}$} \\
 \colhead{} &  \colhead{($M_{\odot}$)} & \colhead{($M_{\odot}$)} & \colhead{($M_{\odot}$ yr$^{-1}$)} & \colhead{} &  \colhead{} &  \colhead{} &  \colhead{} &  \colhead{} &  \colhead{} \\
	\colhead{(1)} & \colhead{(2)} & \colhead{(3)} & \colhead{(4)} & \colhead{(5)} & \colhead{(6)} & \colhead{(7)} & \colhead{(8)} & \colhead{(9)} & \colhead{(10)}}
\startdata
  200803 & 9.87 $\pm$ 0.07 &  10.07 $\pm$ 0.01 &  0.03  $\pm$ 0.01 & 2.36 $\pm$ 0.02 & 0.35 $\pm$ 0.01 &  0.75 $\pm$ 0.02 & 2.84 $\pm$ 0.11 & 0.30 $\pm$ 0.14 &  2.72 $\pm$ 0.16\\
  200805 & 9.77 $\pm$ 0.05 &  9.22  $\pm$ 0.02 & -1.08  $\pm$ 0.10 & 2.67 $\pm$ 0.05 & 0.17 $\pm$ 0.03 &  0.58 $\pm$ 0.03 & 1.80 $\pm$ 0.10 & 0.46 $\pm$ 0.13 &  1.70 $\pm$ 0.15\\
  200845 & 10.02$\pm$ 0.05 &  9.97  $\pm$ 0.01 &  -0.19 $\pm$ 0.02 & 2.70 $\pm$ 0.05 & 0.13 $\pm$ 0.03 &  0.73 $\pm$ 0.03 & 3.39 $\pm$ 0.11 & 0.53 $\pm$ 0.14 &  2.44 $\pm$ 0.16\\
  200847 & 9.66 $\pm$ 0.06 &  9.59  $\pm$ 0.01 &  -0.62 $\pm$ 0.05 & 1.88 $\pm$ 0.08 & 0.64 $\pm$ 0.05 &  1.51 $\pm$ 0.05 & 1.29 $\pm$ 0.16 & 0.49 $\pm$ 0.20 &  2.26 $\pm$ 0.23\\
  200855 & 9.97 $\pm$ 0.06 &  9.64  $\pm$ 0.02 &  -0.45 $\pm$ 0.04 & 2.62 $\pm$ 0.02 & 0.18 $\pm$ 0.01 &  0.51 $\pm$ 0.01 & 2.64 $\pm$ 0.11 & 0.59 $\pm$ 0.14 &  2.23 $\pm$ 0.16\\
  200858 & 9.64 $\pm$ 0.08 &  9.39  $\pm$ 0.02 &  -0.24 $\pm$ 0.03 & 2.49 $\pm$ 0.03 & 0.17 $\pm$ 0.02 &  0.95 $\pm$ 0.02 & 1.76 $\pm$ 0.10 & 0.42 $\pm$ 0.12 &  1.71 $\pm$ 0.13\\
  210063 & 9.33 $\pm$ 0.10 &  9.73  $\pm$ 0.02 &  -0.18 $\pm$ 0.02 & 2.29 $\pm$ 0.01 & 0.19 $\pm$ 0.01 &  0.65 $\pm$ 0.01 & 1.46 $\pm$ 0.09 & 0.65 $\pm$ 0.12 &  1.84 $\pm$ 0.13\\
  210072 & 10.21$\pm$ 0.04 &  10.67 $\pm$ 0.01 &  0.43  $\pm$ 0.01 & 2.66 $\pm$ 0.01 & 0.15 $\pm$ 0.01 &  0.56 $\pm$ 0.01 & 2.96 $\pm$ 0.11 & 0.58 $\pm$ 0.13 &  2.12 $\pm$ 0.15\\
  210073 & 10.19$\pm$ 0.04 &  10.50 $\pm$ 0.01 &  0.44  $\pm$ 0.01 & 2.88 $\pm$ 0.02 & 0.15 $\pm$ 0.01 &  0.60 $\pm$ 0.01 & 2.47 $\pm$ 0.04 & 0.44 $\pm$ 0.04 &  1.79 $\pm$ 0.05\\
  210082 & 8.29 $\pm$ 0.02 &  7.43  $\pm$ 0.01 &  -2.49 $\pm$ 0.04 & 2.20 $\pm$ 0.01 & 0.13 $\pm$ 0.00 &  0.39 $\pm$ 0.01 & 1.19 $\pm$ 0.09 & 0.55 $\pm$ 0.11 &  3.16 $\pm$ 0.13\\
  210084 & 9.71 $\pm$ 0.04 &  9.62  $\pm$ 0.01 &  -0.48 $\pm$ 0.02 & 2.20 $\pm$ 0.01 & 0.19 $\pm$ 0.01 &  0.78 $\pm$ 0.01 & 2.45 $\pm$ 0.15 & 0.05 $\pm$ 0.18 &  0.36 $\pm$ 0.21\\
  210111 & 8.34 $\pm$ 0.02 &  6.99  $\pm$ 0.01 &  -2.27 $\pm$ 0.04 & 2.38 $\pm$ 0.02 & 0.35 $\pm$ 0.01 &  0.71 $\pm$ 0.01 & 2.17 $\pm$ 0.13 & 0.51 $\pm$ 0.16 &  1.05 $\pm$ 0.18\\
  210148 & 9.83 $\pm$ 0.07 &  9.95  $\pm$ 0.01 &  -0.22 $\pm$ 0.03 & 2.60 $\pm$ 0.03 & 0.24 $\pm$ 0.02 &  0.83 $\pm$ 0.02 & 1.86 $\pm$ 0.33 & 0.46 $\pm$ 0.41 &  3.62 $\pm$ 0.47\\
  210171 & 9.68 $\pm$ 0.05 &  10.24 $\pm$ 0.01 &  -0.38 $\pm$ 0.02 & 2.61 $\pm$ 0.01 & 0.10 $\pm$ 0.00 &  0.17 $\pm$ 0.01 & 1.90 $\pm$ 0.08 & 0.43 $\pm$ 0.10 &  2.10 $\pm$ 0.12\\
  210180 & 10.12$\pm$ 0.06 &  10.48 $\pm$ 0.01 &  0.38  $\pm$ 0.01 & 2.52 $\pm$ 0.02 & 0.22 $\pm$ 0.01 &  0.21 $\pm$ 0.01 & 2.20 $\pm$ 0.08 & 0.60 $\pm$ 0.10 &  2.49 $\pm$ 0.12\\
  210354 & 9.98 $\pm$ 0.02 &  9.96  $\pm$ 0.01 &  -0.04 $\pm$ 0.01 & 2.51 $\pm$ 0.01 & 0.11 $\pm$ 0.01 &  0.36 $\pm$ 0.01 & 2.12 $\pm$ 0.02 & 0.33 $\pm$ 0.03 &  1.04 $\pm$ 0.03\\
  210368 & 9.52 $\pm$ 0.03 &  9.01  $\pm$ 0.01 &  -0.74 $\pm$ 0.04 & 2.61 $\pm$ 0.03 & 0.31 $\pm$ 0.01 &  0.53 $\pm$ 0.02 & 4.08 $\pm$ 0.14 & 0.56 $\pm$ 0.18 &  1.90 $\pm$ 0.20\\
  211086 & 9.82 $\pm$ 0.07 &  9.77  $\pm$ 0.01 &  -0.34 $\pm$ 0.03 & 2.92 $\pm$ 0.02 & 0.10 $\pm$ 0.01 &  0.61 $\pm$ 0.02 & 2.85 $\pm$ 0.13 & 0.48 $\pm$ 0.16 &  2.27 $\pm$ 0.18\\
  212984 & 9.79 $\pm$ 0.05 &  9.83  $\pm$ 0.01 &  -0.21 $\pm$ 0.03 & 2.76 $\pm$ 0.02 & 0.18 $\pm$ 0.01 &  0.21 $\pm$ 0.02 & 1.24 $\pm$ 0.34 & 0.01 $\pm$ 0.42 &  0.12 $\pm$ 0.48\\
  212987 & 9.77 $\pm$ 0.04 &  9.10  $\pm$ 0.02 &  -1.06 $\pm$ 0.08 & 2.68 $\pm$ 0.06 & 0.32 $\pm$ 0.03 &  1.33 $\pm$ 0.04 & 2.41 $\pm$ 0.18 & 0.50 $\pm$ 0.22 &  2.82 $\pm$ 0.25\\
  212988 & 8.86 $\pm$ 0.12 &  8.63  $\pm$ 0.02 &  -1.3  $\pm$ 0.07 & 3.30 $\pm$ 0.04 & 0.26 $\pm$ 0.02 &  1.22 $\pm$ 0.02 & 1.92 $\pm$ 0.23 & 0.45 $\pm$ 0.28 &  2.80 $\pm$ 0.32\\
  212989 & 9.82 $\pm$ 0.09 &  10.02 $\pm$ 0.02 &  0.00  $\pm$ 0.02 & 2.37 $\pm$ 0.02 & 0.35 $\pm$ 0.01 &  0.33 $\pm$ 0.01 & 2.21 $\pm$ 0.07 & 0.58 $\pm$ 0.09 &  2.03 $\pm$ 0.11\\
  212990 & 9.99 $\pm$ 0.02 &  9.26  $\pm$ 0.02 &  -0.82 $\pm$ 0.05 & 2.51 $\pm$ 0.03 & 0.25 $\pm$ 0.02 &  0.65 $\pm$ 0.02 & 1.85 $\pm$ 0.29 & 0.54 $\pm$ 0.36 &  0.67 $\pm$ 0.41\\
  212994 & 9.71 $\pm$ 0.10 &  10.07 $\pm$ 0.01 &  0.10  $\pm$ 0.02 & 2.66 $\pm$ 0.01 & 0.14 $\pm$ 0.01 &  0.26 $\pm$ 0.01 & 2.10 $\pm$ 0.05 & 0.48 $\pm$ 0.07 &  2.27 $\pm$ 0.08\\
  212996 & 9.81 $\pm$ 0.06 &  9.37  $\pm$ 0.02 &  -0.45 $\pm$ 0.02 & 2.88 $\pm$ 0.02 & 0.32 $\pm$ 0.01 &  0.53 $\pm$ 0.01 & 2.48 $\pm$ 0.07 & 0.37 $\pm$ 0.08 &  1.60 $\pm$ 0.09\\
  213010 & 9.00 $\pm$ 0.04 &   8.43 $\pm$ 0.02 &  -2.03 $\pm$ 0.32 & 2.27 $\pm$ 0.03 & 0.28 $\pm$ 0.02 &  0.42 $\pm$ 0.02 & 1.47 $\pm$ 0.31 & 0.44 $\pm$ 0.44 &  2.98 $\pm$ 0.42\\
  213058 & 9.51 $\pm$ 0.09 &  10.01 $\pm$ 0.01 &  -0.15 $\pm$ 0.01 & 2.65 $\pm$ 0.01 & 0.16 $\pm$ 0.01 &  0.21 $\pm$ 0.01 & 2.10 $\pm$ 0.04 & 0.52 $\pm$ 0.05 &  2.01 $\pm$ 0.06\\
  213062 & 9.27 $\pm$ 0.03 &  8.89  $\pm$ 0.01 &  -1.17 $\pm$ 0.06 & 2.74 $\pm$ 0.04 & 0.40 $\pm$ 0.02 &  1.20 $\pm$ 0.02 & 2.55 $\pm$ 0.17 & 0.55 $\pm$ 0.21 &  1.16 $\pm$ 0.24\\
  213064 & 8.42 $\pm$ 0.01 &  7.76  $\pm$ 0.01 &  -1.93 $\pm$ 0.01 & 2.82 $\pm$ 0.01 & 0.30 $\pm$ 0.01 &  0.08 $\pm$ 0.01 & 2.84 $\pm$ 0.01 & 0.40 $\pm$ 0.01 &  0.19 $\pm$ 0.01\\
  213069 & 9.36 $\pm$ 0.04 &  8.56  $\pm$ 0.02 &  -1.50 $\pm$ 0.35 & 2.64 $\pm$ 0.04 & 0.14 $\pm$ 0.02 &  0.58 $\pm$ 0.02 & 1.62 $\pm$ 0.31 & 0.69 $\pm$ 0.48 &  3.88 $\pm$ 0.41\\
  213072 & 9.58 $\pm$ 0.03 &  9.82  $\pm$ 0.01 &  -0.16 $\pm$ 0.01 & 2.29 $\pm$ 0.02 & 0.09 $\pm$ 0.01 &  0.39 $\pm$ 0.01 & 2.23 $\pm$ 0.06 & 0.61 $\pm$ 0.07 &  2.22 $\pm$ 0.09\\
  213074 & 7.98 $\pm$ 0.01 &  6.42  $\pm$ 0.03 &  -2.33 $\pm$ 0.08 & 3.56 $\pm$ 0.04 & 0.54 $\pm$ 0.02 &  0.94 $\pm$ 0.03 & 2.29 $\pm$ 0.06 & 0.50 $\pm$ 0.08 &  2.10 $\pm$ 0.09\\
  213076 & 9.81 $\pm$ 0.04 &  9.31  $\pm$ 0.01 &  -0.93 $\pm$ 0.06 & 2.42 $\pm$ 0.03 & 0.03 $\pm$ 0.01 &  0.52 $\pm$ 0.02 & 2.22 $\pm$ 0.31 & 0.33 $\pm$ 0.38 &  2.97 $\pm$ 0.44\\
  213083 & 9.45 $\pm$ 0.08 &  9.31  $\pm$ 0.01 &  -0.76 $\pm$ 0.05 & 2.60 $\pm$ 0.03 & 0.22 $\pm$ 0.02 &  0.24 $\pm$ 0.02 & 2.45 $\pm$ 0.22 & 0.64 $\pm$ 0.27 &  2.21 $\pm$ 0.31\\
  213085 & 9.75 $\pm$ 0.04 &  8.96  $\pm$ 0.02 &  -1.14 $\pm$ 0.24 & 2.72 $\pm$ 0.03 & 0.34 $\pm$ 0.02 &  1.30 $\pm$ 0.02 & 0.93 $\pm$ 0.34 & 0.47 $\pm$ 0.42 &  3.44 $\pm$ 0.28\\
  213087 & 9.77 $\pm$ 0.05 &  9.62  $\pm$ 0.01 &  -0.53 $\pm$ 0.04 & 3.29 $\pm$ 0.03 & 0.20 $\pm$ 0.02 &  0.42 $\pm$ 0.02 & 4.43 $\pm$ 0.15 & 0.40 $\pm$ 0.18 &  2.54 $\pm$ 0.21\\
  213090 & 9.26 $\pm$ 0.03 &  8.56  $\pm$ 0.02 &  -0.71 $\pm$ 0.02 & 1.99 $\pm$ 0.02 & 0.68 $\pm$ 0.01 &  0.92 $\pm$ 0.01 & 1.17 $\pm$ 0.03 & 0.90 $\pm$ 0.04 &  1.53 $\pm$ 0.05\\
  213092 & 9.45 $\pm$ 0.04 &  9.20  $\pm$ 0.02 &  -0.60 $\pm$ 0.02 & 2.56 $\pm$ 0.01 & 0.38 $\pm$ 0.01 &  0.28 $\pm$ 0.01 & 2.21 $\pm$ 0.03 & 0.55 $\pm$ 0.04 &  1.46 $\pm$ 0.04\\
  213198 & 9.68 $\pm$ 0.07 &  9.94  $\pm$ 0.01 &  -0.38 $\pm$ 0.04 & 2.26 $\pm$ 0.03 & 0.20 $\pm$ 0.02 &  0.66 $\pm$ 0.02 & 1.85 $\pm$ 0.07 & 0.53 $\pm$ 0.08 &  2.29 $\pm$ 0.10\\
  213573 & 9.76 $\pm$ 0.07 &  9.38  $\pm$ 0.02 &  -0.83 $\pm$ 0.08 & 2.80 $\pm$ 0.04 & 0.12 $\pm$ 0.02 &  0.65 $\pm$ 0.03 & 2.64 $\pm$ 0.13 & 0.55 $\pm$ 0.15 &  0.50 $\pm$ 0.18\\
  213574 & 9.21 $\pm$ 0.10 &  9.75  $\pm$ 0.01 &  -0.44 $\pm$ 0.04 & 3.04 $\pm$ 0.02 & 0.19 $\pm$ 0.01 &  0.48 $\pm$ 0.01 & 2.38 $\pm$ 0.09 & 0.42 $\pm$ 0.11 &  1.90 $\pm$ 0.13\\
  213583 & 9.72 $\pm$ 0.06 &  9.90  $\pm$ 0.01 &  -0.16 $\pm$ 0.03 & 2.58 $\pm$ 0.06 & 0.19 $\pm$ 0.03 &  0.90 $\pm$ 0.04 & 2.70 $\pm$ 0.10 & 0.28 $\pm$ 0.12 &  2.21 $\pm$ 0.14\\
  213586 & 9.68 $\pm$ 0.08 &  10.69 $\pm$ 0.01 &  -0.39 $\pm$ 0.03 & 3.68 $\pm$ 0.01 & 0.17 $\pm$ 0.01 &  0.35 $\pm$ 0.01 & 4.41 $\pm$ 0.05 & 0.09 $\pm$ 0.06 &  0.06 $\pm$ 0.06\\
  213588 & 9.79 $\pm$ 0.07 &  9.18  $\pm$ 0.02 &  -0.77 $\pm$ 0.16 & 2.92 $\pm$ 0.06 & 0.38 $\pm$ 0.03 &  0.81 $\pm$ 0.04 & 2.63 $\pm$ 0.29 & 0.29 $\pm$ 0.36 &  1.58 $\pm$ 0.41\\
  213590 & 10.09$\pm$ 0.04 &  9.46  $\pm$ 0.02 &  -0.53 $\pm$ 0.04 & 2.01 $\pm$ 0.02 & 0.43 $\pm$ 0.01 &  0.38 $\pm$ 0.02 & 1.56 $\pm$ 0.05 & 0.01 $\pm$ 0.06 &  1.00 $\pm$ 0.06\\
  213596 & 9.95 $\pm$ 0.07 &  9.95  $\pm$ 0.01 &  -0.19 $\pm$ 0.03 & 2.63 $\pm$ 0.02 & 0.11 $\pm$ 0.01 &  0.23 $\pm$ 0.01 & 2.02 $\pm$ 0.73 & 0.58 $\pm$ 0.40 &  3.56 $\pm$ 0.44\\
  213603 & 9.53 $\pm$ 0.09 &  9.42  $\pm$ 0.02 &  -0.47 $\pm$ 0.06 & 2.69 $\pm$ 0.04 & 0.43 $\pm$ 0.02 &  0.55 $\pm$ 0.03 & 2.48 $\pm$ 0.09 & 0.42 $\pm$ 0.11 &  1.42 $\pm$ 0.12\\
  213609 & 10.00$\pm$ 0.06 &  9.86  $\pm$ 0.02 &  -0.02 $\pm$ 0.03 & 2.67 $\pm$ 0.02 & 0.14 $\pm$ 0.01 &  0.21 $\pm$ 0.01 & 2.55 $\pm$ 0.06 & 0.27 $\pm$ 0.08 &  1.60 $\pm$ 0.09\\
  213611 & 9.95 $\pm$ 0.04 &  10.44 $\pm$ 0.01 &  -0.23 $\pm$ 0.02 & 3.18 $\pm$ 0.02 & 0.08 $\pm$ 0.01 &  0.28 $\pm$ 0.01 & 3.72 $\pm$ 0.07 & 0.41 $\pm$ 0.09 &  2.00 $\pm$ 0.10\\
  215236 & 9.76 $\pm$ 0.06 &  8.67  $\pm$ 0.03 &  -1.23 $\pm$ 0.18 & 2.27 $\pm$ 0.15 & 0.05 $\pm$ 0.08 &  1.15 $\pm$ 0.09 & 2.19 $\pm$ 0.16 & 0.64 $\pm$ 0.20 &  2.73 $\pm$ 0.23\\
  215237 & 9.57 $\pm$ 0.07 &  8.45  $\pm$ 0.04 &  -1.24 $\pm$ 0.13 & 2.64 $\pm$ 0.17 & 0.26 $\pm$ 0.10 &  2.07 $\pm$ 0.11 & 1.44 $\pm$ 0.28 & 0.04 $\pm$ 0.35 &  0.53 $\pm$ 0.40\\
  215239 & 9.86 $\pm$ 0.05 &  8.71  $\pm$ 0.04 &  -1.05 $\pm$ 0.56 & 2.36 $\pm$ 0.09 & 0.34 $\pm$ 0.05 &  0.81 $\pm$ 0.06 & 1.89 $\pm$ 0.13 & 0.03 $\pm$ 0.16 &  0.24 $\pm$ 0.19\\
  215240 & 7.52 $\pm$ 0.06 &  6.43  $\pm$ 0.04 &  -3.18 $\pm$ 1.03 & 2.33 $\pm$ 0.10 & 0.30 $\pm$ 0.06 &  0.67 $\pm$ 0.06 & 1.74 $\pm$ 0.56 & 0.13 $\pm$ 0.64 &  2.58 $\pm$ 0.38\\
  215241 & 8.15 $\pm$ 0.03 &  6.68  $\pm$ 0.03 &  -3.16 $\pm$ 1.92 & 2.61 $\pm$ 0.05 & 0.28 $\pm$ 0.03 &  0.64 $\pm$ 0.03 & 0.34 $\pm$ 0.41 & 0.48 $\pm$ 0.57 &  2.56 $\pm$ 0.31\\
  215242 & 9.52 $\pm$ 0.06 &  8.56  $\pm$ 0.03 &  -1.16 $\pm$ 0.23 & 2.85 $\pm$ 0.09 & 0.45 $\pm$ 0.05 &  1.25 $\pm$ 0.06 & 1.79 $\pm$ 0.31 & 0.57 $\pm$ 0.54 &  3.04 $\pm$ 0.36\\
  215761 & 9.89 $\pm$ 0.05 &  9.50  $\pm$ 0.01 &  -0.66 $\pm$ 0.04 & 2.90 $\pm$ 0.02 & 0.24 $\pm$ 0.01 &  0.14 $\pm$ 0.02 & 3.30 $\pm$ 0.04 & 0.45 $\pm$ 0.05 &  0.99 $\pm$ 0.06\\
  219116 & 9.52 $\pm$ 0.05 &  8.42  $\pm$ 0.03 &  -0.34 $\pm$ 0.06 & 2.31 $\pm$ 0.16 & 0.49 $\pm$ 0.09 &  1.45 $\pm$ 0.10 & 2.11 $\pm$ 0.08 & 0.58 $\pm$ 0.10 &  2.37 $\pm$ 0.12\\
  6086   & 10.03$\pm$ 0.06 &  9.95  $\pm$ 0.01 &  -0.40 $\pm$ 0.02 & 3.22 $\pm$ 0.02 & 0.24 $\pm$ 0.01 &  0.69 $\pm$ 0.02 & 1.92 $\pm$ 0.12 & 0.61 $\pm$ 0.14 &  2.56 $\pm$ 0.17\\
  6091   & 10.31$\pm$ 0.03 &  10.46 $\pm$ 0.01 &  0.08  $\pm$ 0.01 & 3.86 $\pm$ 0.01 & 0.32 $\pm$ 0.01 &  0.53 $\pm$ 0.01 & 2.76 $\pm$ 0.02 & 0.53 $\pm$ 0.02 &  1.63 $\pm$ 0.02\\
  6093   & 9.91 $\pm$ 0.05 &  10.71 $\pm$ 0.01 &  -0.17 $\pm$ 0.02 & 3.37 $\pm$ 0.01 & 0.17 $\pm$ 0.01 &  0.52 $\pm$ 0.01 & 2.08 $\pm$ 0.06 & 0.42 $\pm$ 0.07 &  0.25 $\pm$ 0.08\\
  6209   & 8.82 $\pm$ 0.01 &  8.64  $\pm$ 0.01 &  -1.01 $\pm$ 0.01 & 2.61 $\pm$ 0.01 & 0.19 $\pm$ 0.00 &  0.44 $\pm$ 0.00 & 2.52 $\pm$ 0.01 & 0.79 $\pm$ 0.00 &  1.29 $\pm$ 0.01\\
  6245   & 8.00 $\pm$ 0.04 &  9.13  $\pm$ 0.01 &  -0.56 $\pm$ 0.01 & 3.27 $\pm$ 0.01 & 0.22 $\pm$ 0.00 &  0.30 $\pm$ 0.00 & 2.07 $\pm$ 0.01 & 0.89 $\pm$ 0.01 &  1.92 $\pm$ 0.01\\
  6248   & 8.26 $\pm$ 0.01 &  6.80  $\pm$ 0.02 &  -2.02 $\pm$ 0.06 & 2.73 $\pm$ 0.04 & 0.36 $\pm$ 0.02 &  0.19 $\pm$ 0.03 & 1.93 $\pm$ 0.19 & 0.51 $\pm$ 0.24 &  2.94 $\pm$ 0.27\\
  6288   & 9.66 $\pm$ 0.03 &  9.19  $\pm$ 0.01 &  -0.83 $\pm$ 0.02 & 2.60 $\pm$ 0.01 & 0.12 $\pm$ 0.01 &  0.07 $\pm$ 0.01 & 2.14 $\pm$ 0.02 & 0.37 $\pm$ 0.02 &  0.85 $\pm$ 0.03\\
  6438   & 8.58 $\pm$ 0.01 &  8.52  $\pm$ 0.01 &  -1.78 $\pm$ 0.01 & 3.24 $\pm$ 0.01 & 0.13 $\pm$ 0.01 &  0.04 $\pm$ 0.01 & 1.32 $\pm$ 0.05 & 0.99 $\pm$ 0.07 &  2.38 $\pm$ 0.08\\
  6470   & 10.09$\pm$ 0.02 &  10.55 $\pm$ 0.01 &  0.60  $\pm$ 0.01 & 3.20 $\pm$ 0.01 & 0.37 $\pm$ 0.01 &  0.44 $\pm$ 0.01 & 3.52 $\pm$ 0.01 & 0.58 $\pm$ 0.02 &  1.09 $\pm$ 0.02\\
  6474   & 9.50 $\pm$ 0.01 &  9.78  $\pm$ 0.01 &  -0.84 $\pm$ 0.01 & 3.68 $\pm$ 0.01 & 0.15 $\pm$ 0.00 &  0.95 $\pm$ 0.00 & 3.23 $\pm$ 0.01 & 0.34 $\pm$ 0.01 &  1.36 $\pm$ 0.01\\
  6475   & 10.20$\pm$ 0.01 &  10.36 $\pm$ 0.01 &  -0.01 $\pm$ 0.01 & 3.52 $\pm$ 0.01 & 0.16 $\pm$ 0.01 &  1.59 $\pm$ 0.01 & 3.09 $\pm$ 0.01 & 0.52 $\pm$ 0.02 &  2.02 $\pm$ 0.02\\
  6477   & 10.08$\pm$ 0.01 &  9.92  $\pm$ 0.01 &  -0.02 $\pm$ 0.01 & 2.61 $\pm$ 0.01 & 0.26 $\pm$ 0.00 &  0.23 $\pm$ 0.01 & 1.99 $\pm$ 0.01 & 0.61 $\pm$ 0.01 &  1.32 $\pm$ 0.01\\
  6482   & 9.62 $\pm$ 0.05 &  10.87 $\pm$ 0.01 &  0.95  $\pm$ 0.01 & 3.20 $\pm$ 0.01 & 0.25 $\pm$ 0.01 &  0.35 $\pm$ 0.00 & 1.98 $\pm$ 0.01 & 0.47 $\pm$ 0.01 &  0.95 $\pm$ 0.01\\
\enddata
\end{deluxetable*}

\renewcommand\thefigure{\thesection.\arabic{figure}}
\begin{figure*}
	\centering
	\includegraphics[width=0.8\hsize]{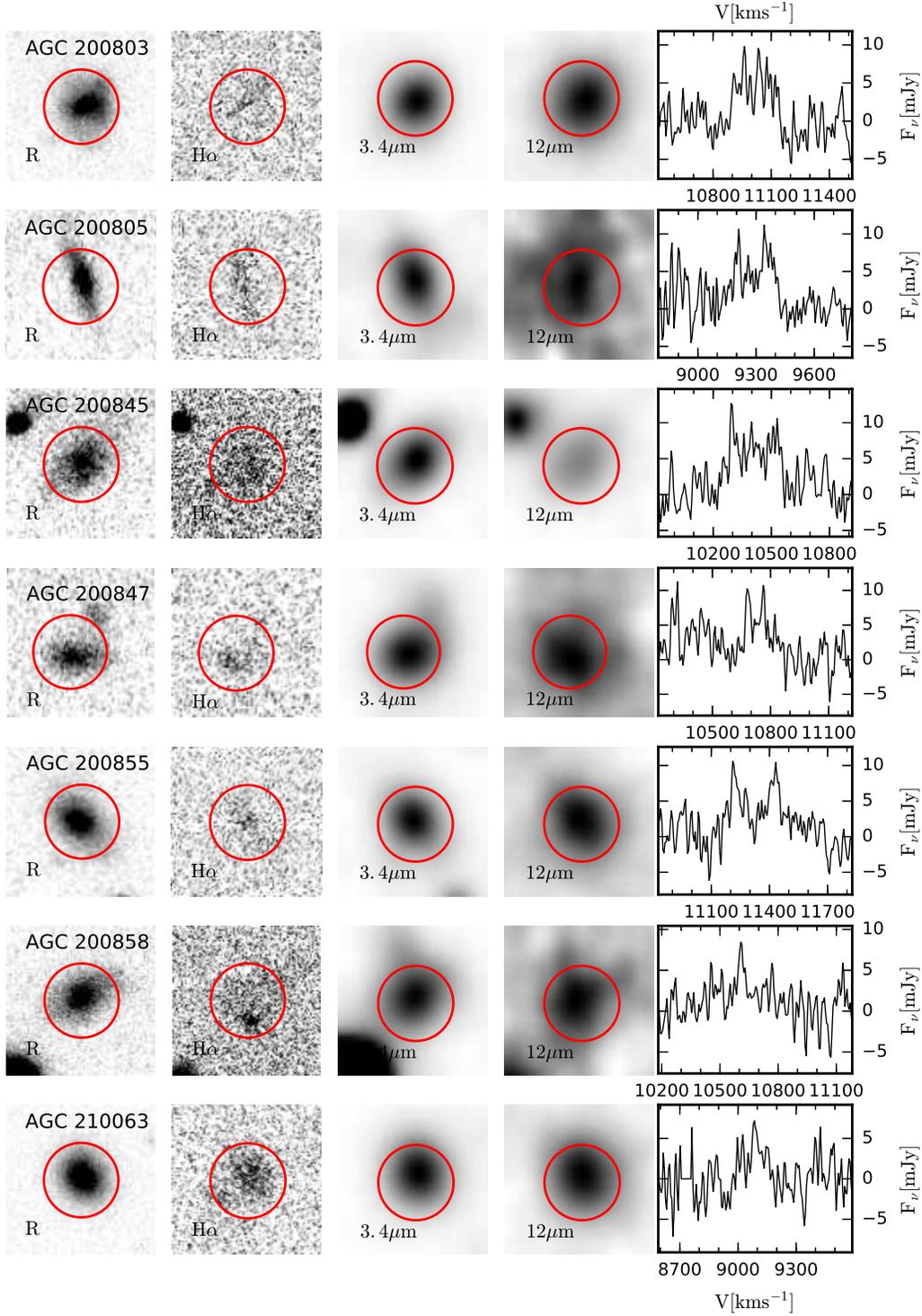}
	\setcounter{figure}{0}
	\caption{Images in broadband $R$, continuum-subtracted narrowband H$\alpha$, WISE IR 3.4$\mu$m and 12$\mu$m, along with the Global \hi\ profiles from ALFALFA for all 70 targets. Petrosian radii of SDSS $r$ band are overlaid on the images.}
	\label{fig_imageshow}
\end{figure*}
\begin{figure*}
	\centering
	\includegraphics[width=0.8\hsize]{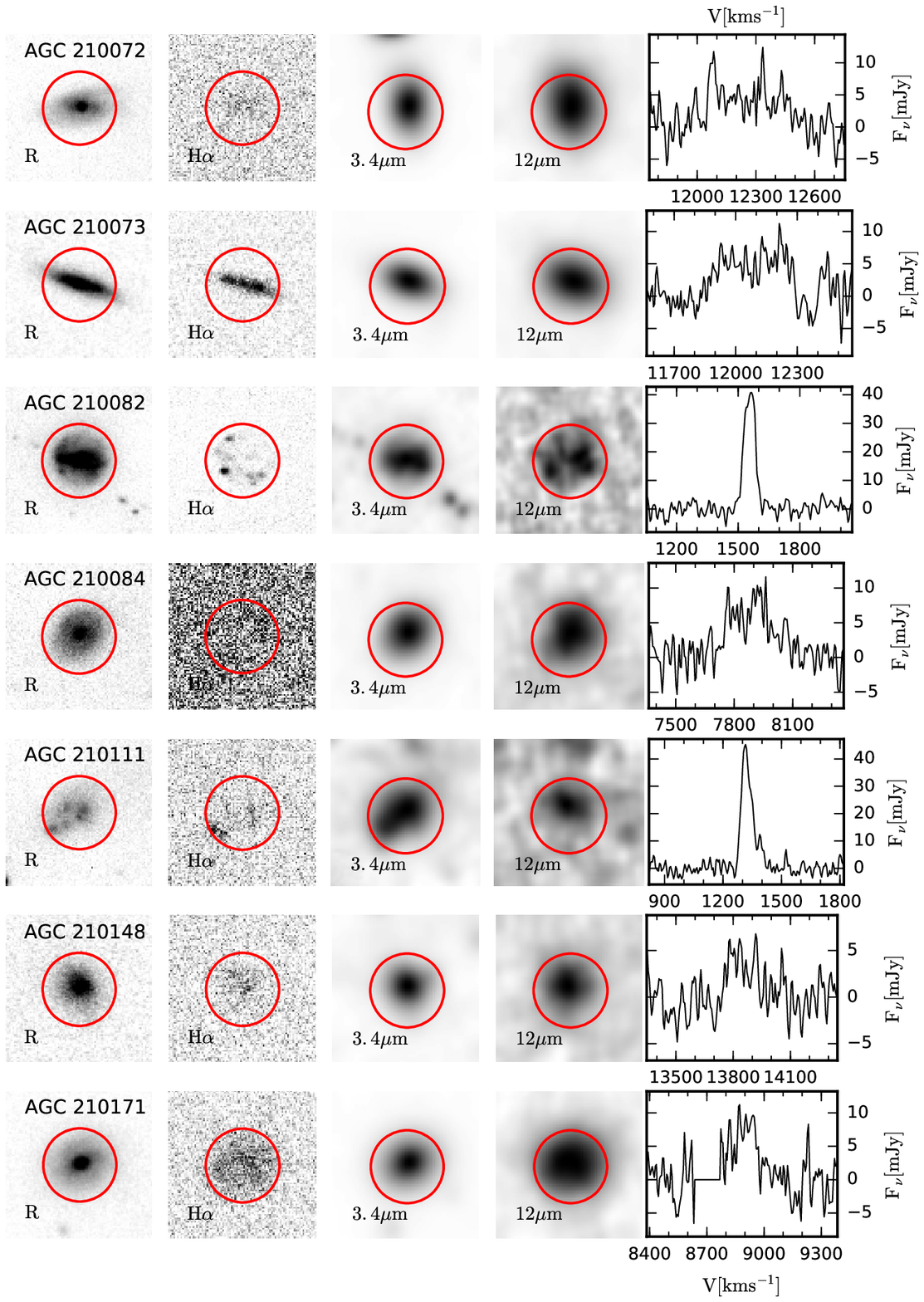}
	\setcounter{figure}{0}
	\caption{(Continued)}
\end{figure*}
\begin{figure*}
	\centering
	\includegraphics[width=0.8\hsize]{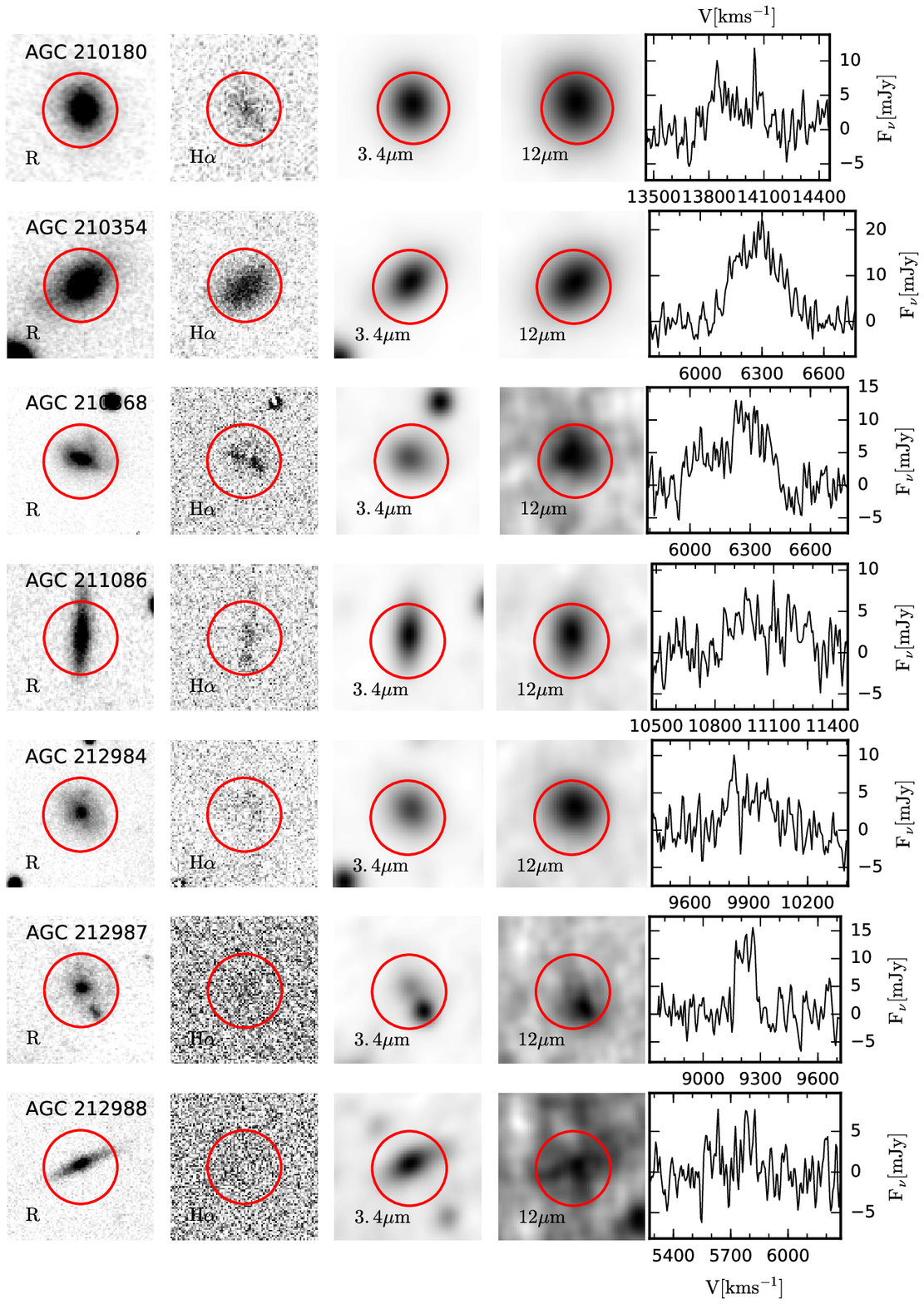}
	\setcounter{figure}{0}
	\caption{(Continued)}
\end{figure*}\begin{figure*}
	\centering
	\includegraphics[width=0.8\hsize]{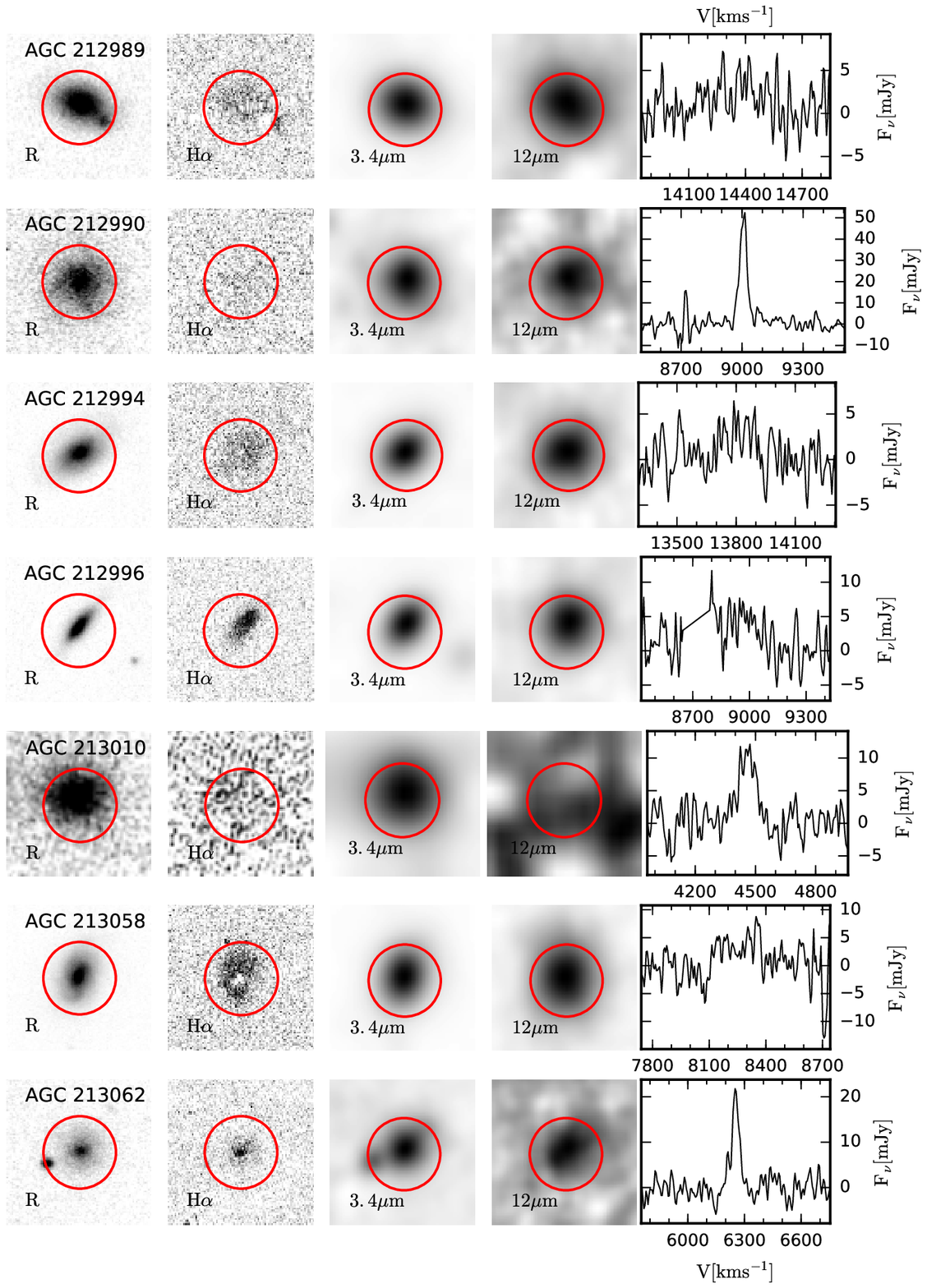}
	\setcounter{figure}{0}
	\caption{(Continued)}
\end{figure*}\begin{figure*}
	\centering
	\includegraphics[width=0.8\hsize]{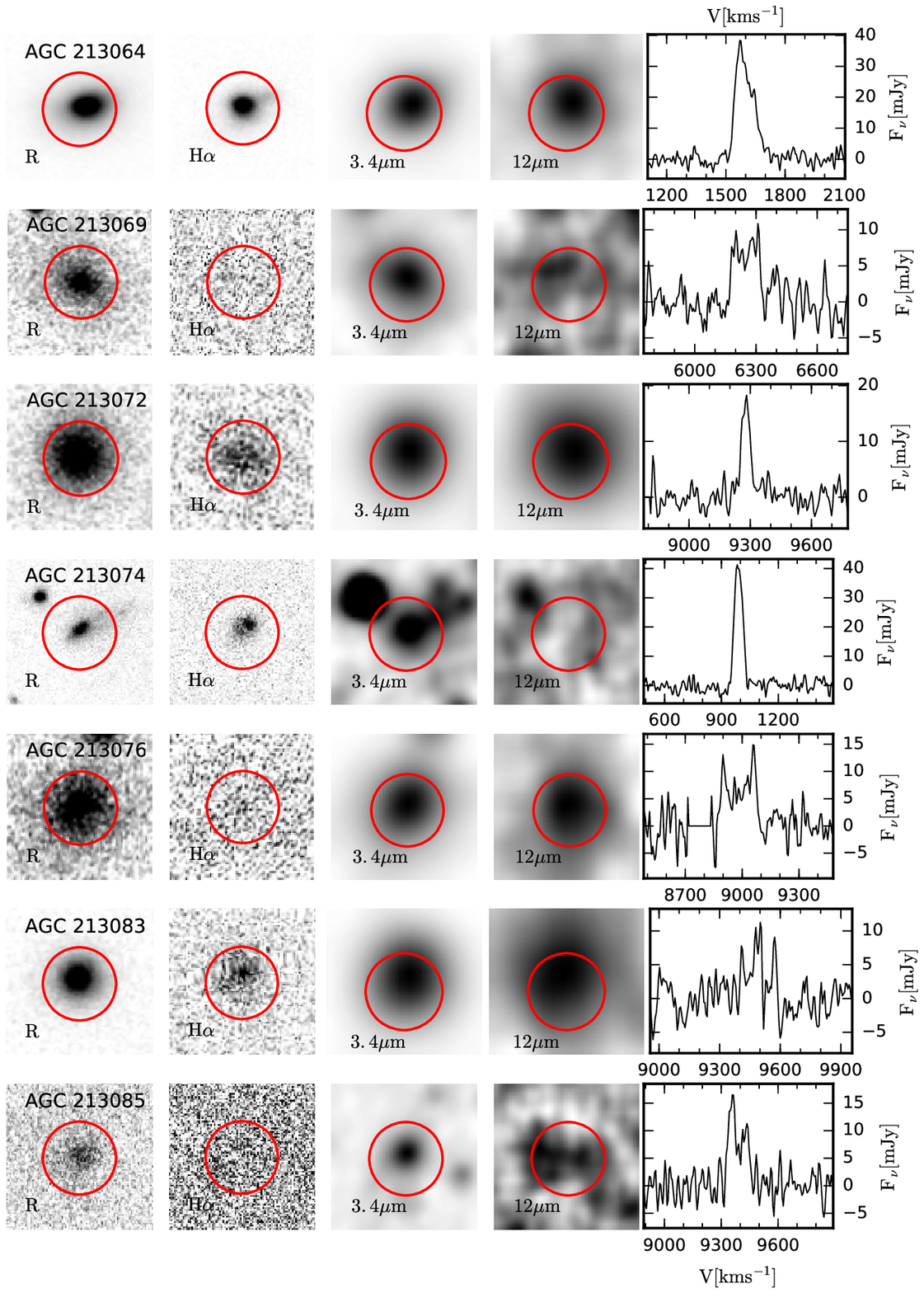}
	\setcounter{figure}{0}
	\caption{(Continued)}
\end{figure*}\begin{figure*}
	\centering
	\includegraphics[width=0.8\hsize]{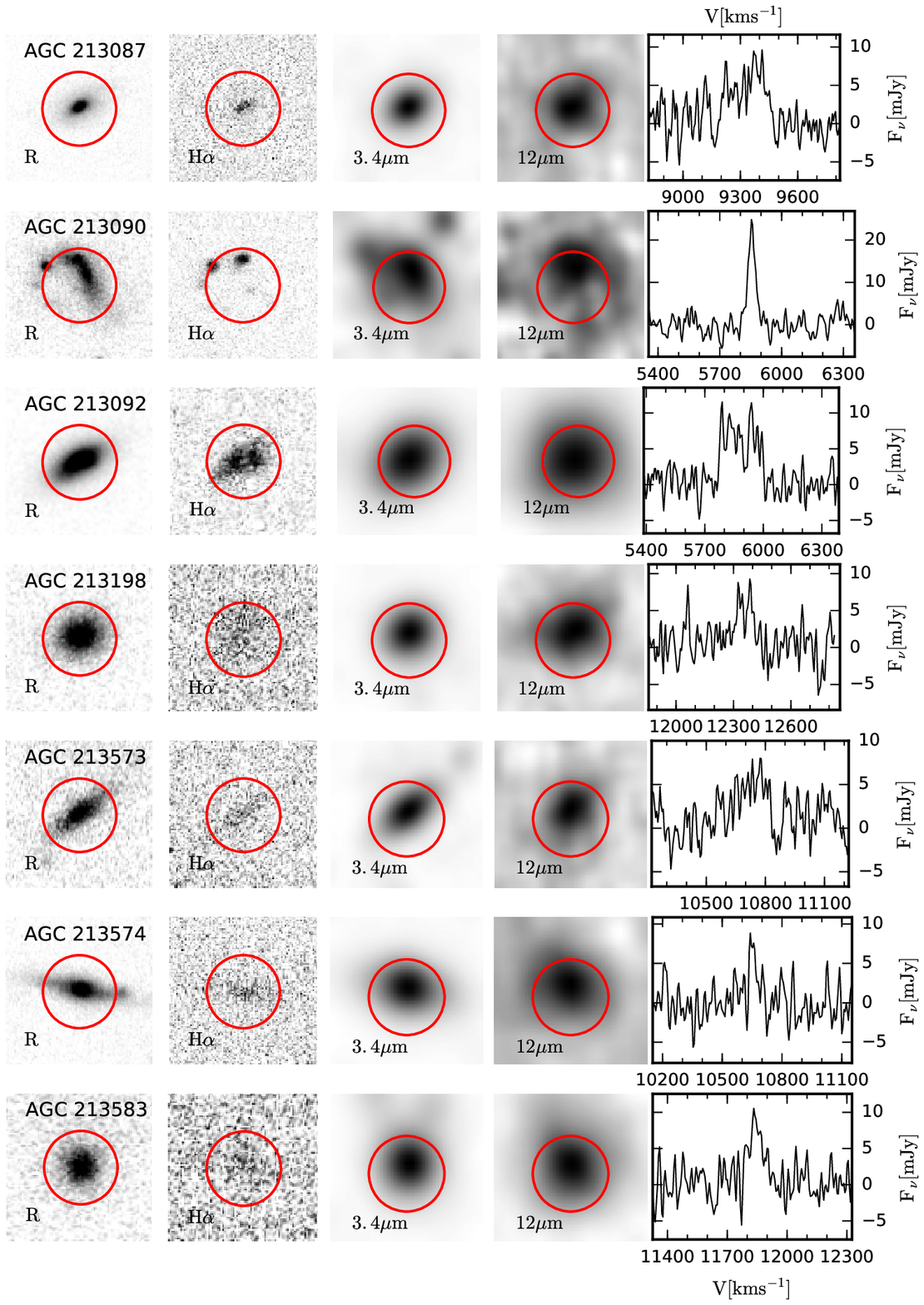}
	\setcounter{figure}{0}
	\caption{(Continued)}
\end{figure*}\begin{figure*}
	\centering
	\includegraphics[width=0.8\hsize]{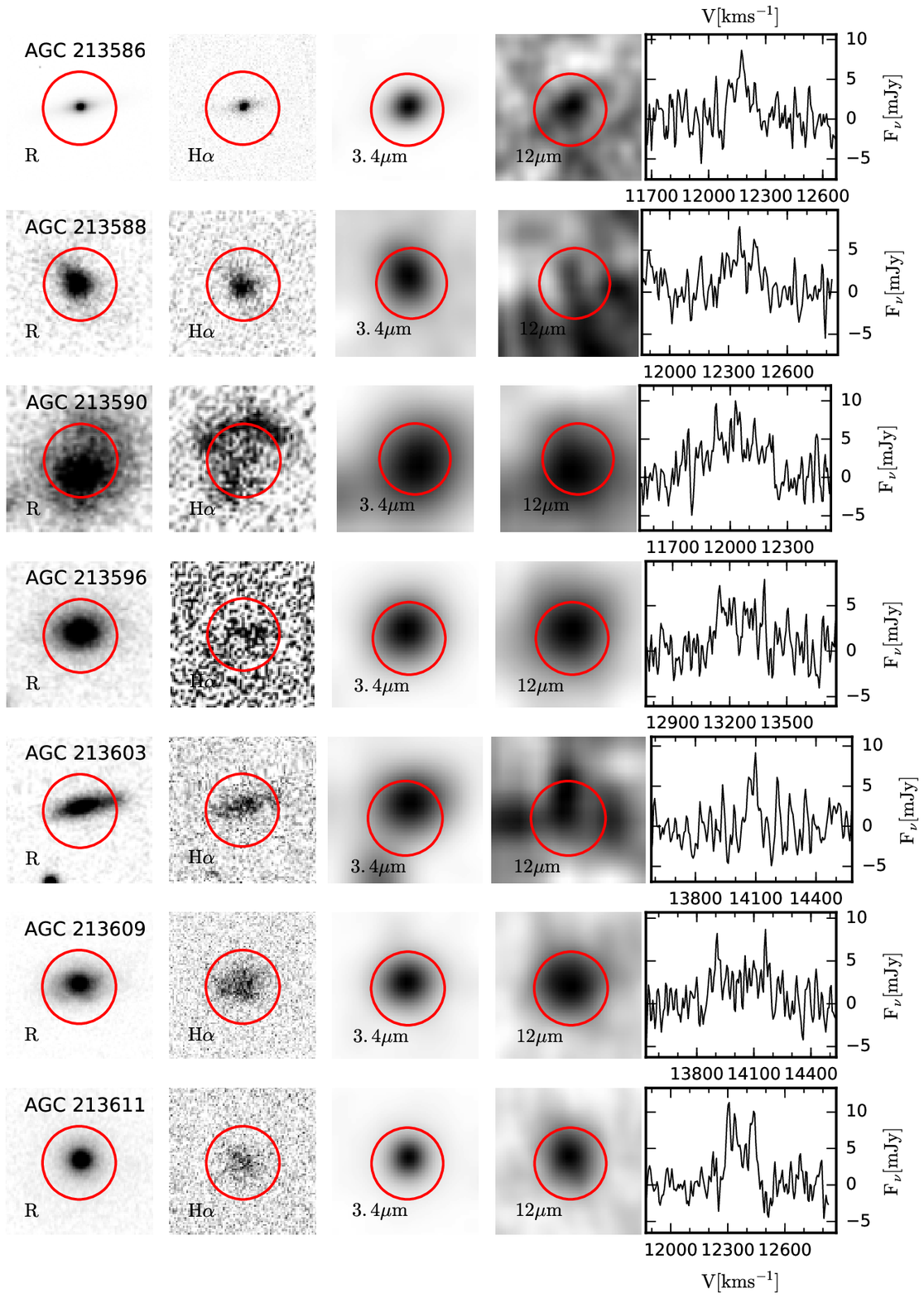}
	\setcounter{figure}{0}
	\caption{(Continued)}
\end{figure*}\begin{figure*}
	\centering
	\includegraphics[width=0.8\hsize]{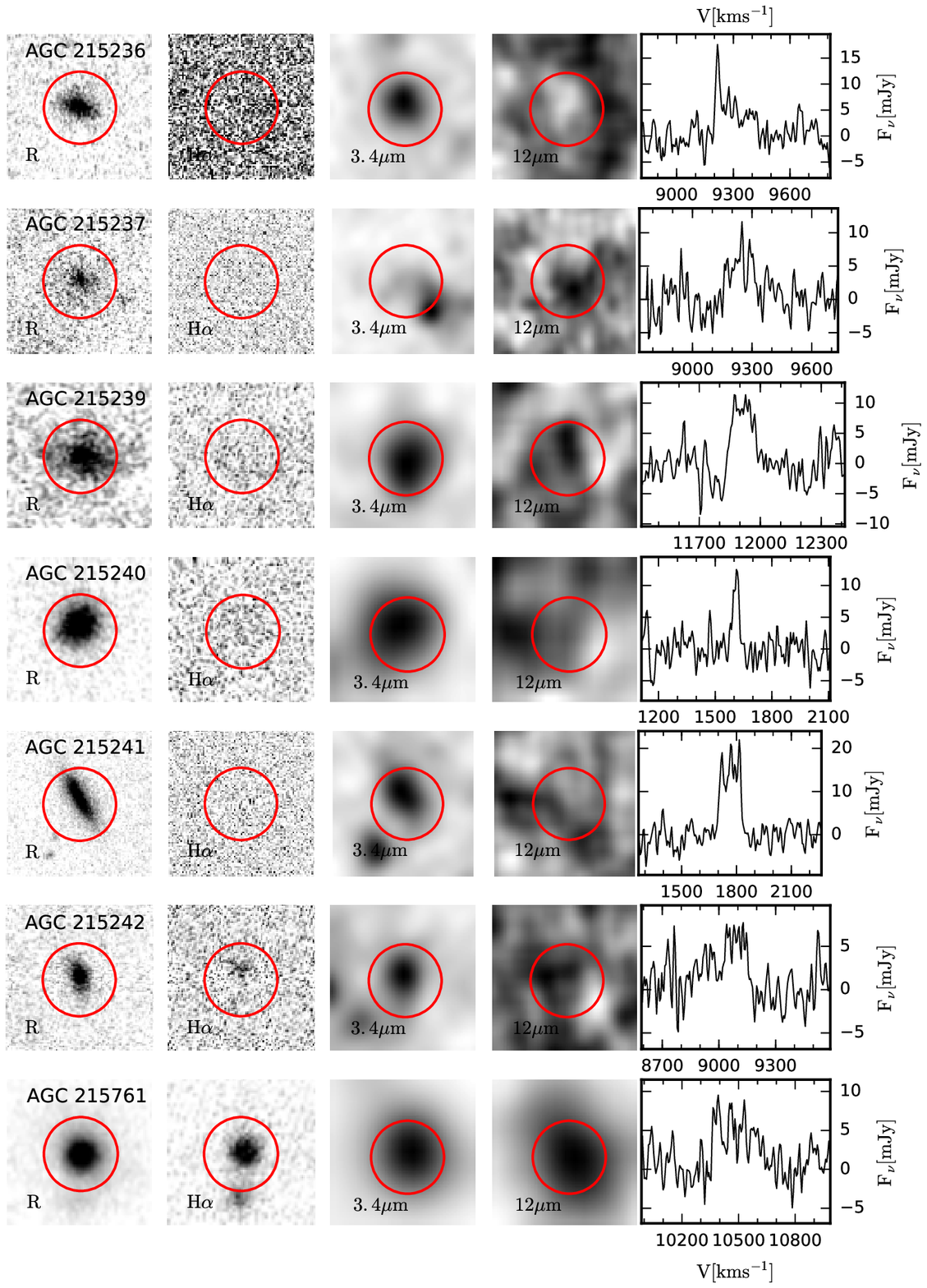}
	\setcounter{figure}{0}
	\caption{(Continued)}
\end{figure*}\begin{figure*}
	\centering
	\includegraphics[width=0.8\hsize]{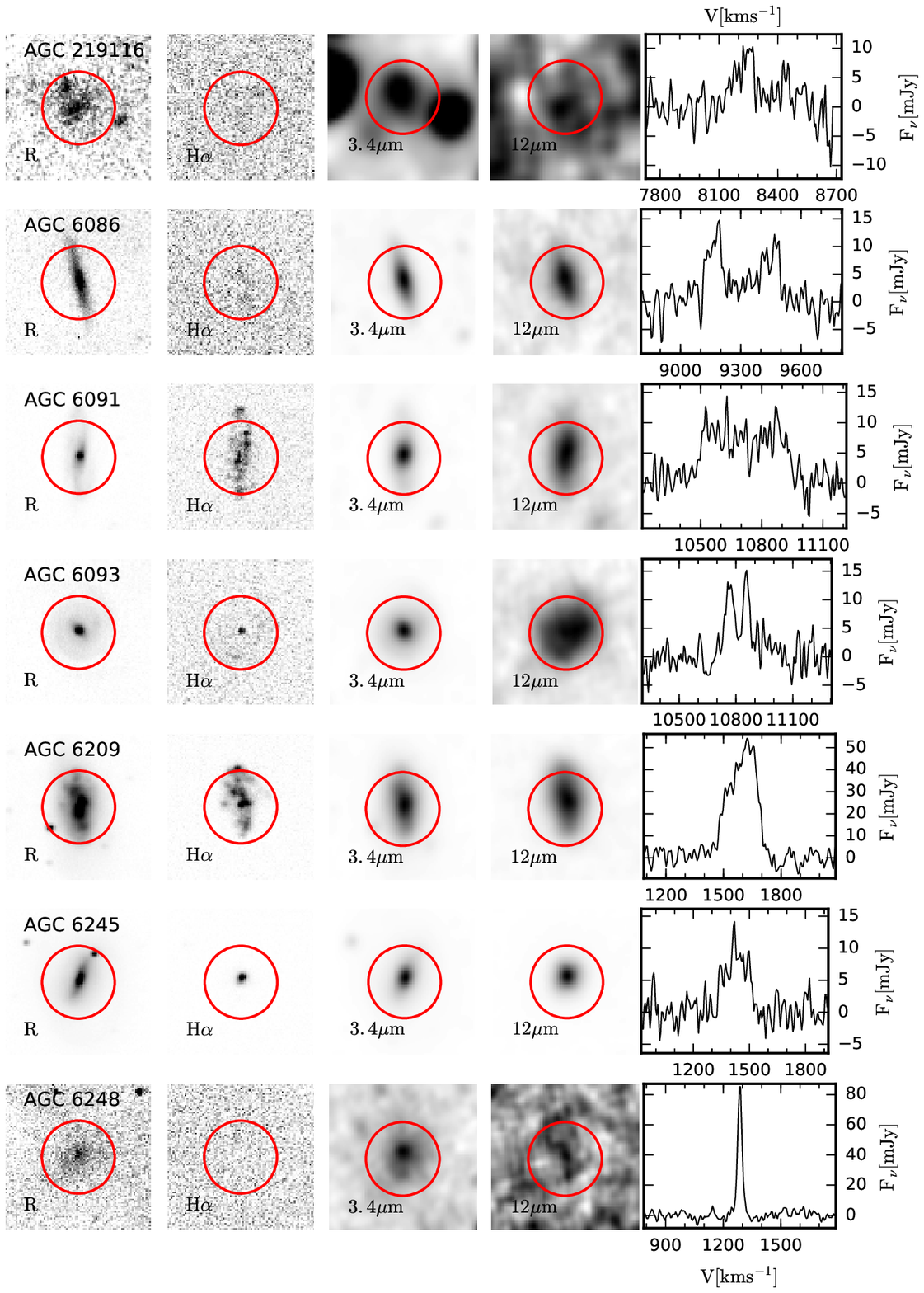}
	\setcounter{figure}{0}
	\caption{(Continued)}
\end{figure*}\begin{figure*}
	\centering
	\includegraphics[width=0.8\hsize]{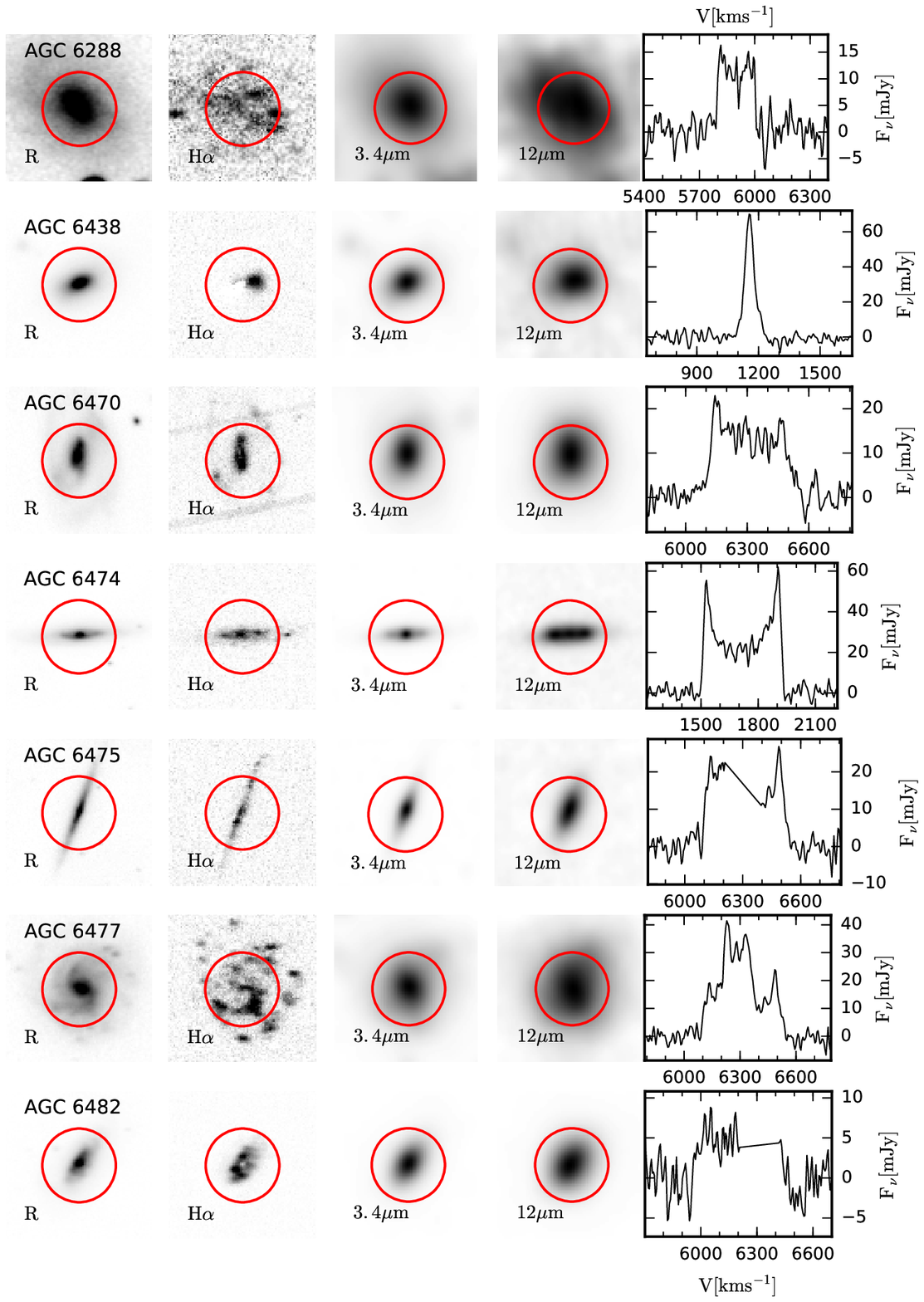}
	\setcounter{figure}{0}
	\caption{(Continued)}
\end{figure*}

\end{CJK*}
\end{document}